\documentclass[12pt]{article}

\usepackage{amsmath,amssymb,amsbsy,amsfonts,latexsym,graphicx}
\usepackage{hyperref}
\usepackage{color, array, subfigure}
\usepackage{cite}
\usepackage[title]{appendix}

\allowdisplaybreaks

\evensidemargin \oddsidemargin

\begin{document}

\begin{titlepage}

\renewcommand{\thefootnote}{\fnsymbol{footnote}}



\vspace{15mm}
\baselineskip 9mm
\begin{center}
  {\Large \bf 
{  
Impurity-Driven Metal-Insulator Transitions in Holography} 
 }
\end{center}

\baselineskip 6mm
\vspace{10mm}
\begin{center}
Yunseok Seo$^{1,a}$, Youngjun Ahn$^{2,b}$, Keun-Young Kim$^{2,3,c}$, Sang-Jin Sin$^{4,d}$ and Kyung Kiu Kim$^{1,e}$
 \\[10mm] 
  $^1${\sl College of General Education, Kookmin University, Seoul 02707, Korea}
   \\[3mm]
  $^2${\sl Department of Physics and Photon Science, Gwangju Institute of Science and Technology, \\
123 Cheomdan-gwagiro, Gwangju 61005, Korea}
    \\[3mm]
    $^3${\sl Research Center for Photon Science Technology, Gwangju Institute of Science and Technology, \\
123 Cheomdan-gwagiro, Gwangju 61005, Korea}
    \\[3mm] 
$^4${\sl Department of Physics, Hanyang University, Seoul 04764, Korea}
    \\[3mm]      
  {\tt  ${}^a$yseo@kookmin.ac.kr, ${}^b$yongjunahn619@gmail.com, ${}^c$fortoe@gist.ac.kr, ${}^d$sjsin@hanyang.ac.kr,${}^e$kimkyungkiu@kookmin.ac.kr}
\end{center}

\thispagestyle{empty}

\vspace{1cm}
\begin{center}
{\bf Abstract}
\end{center}
\noindent
In this work, we study Metal-Insulator transition in a holographic model containing an interaction between the order parameter and charge-carrier density. It turns out that the impurity density of this model can drive the phase transition whose ordered phase corresponds to the insulating phase. The temperature behavior of DC conductivity distinguishes the insulating phase from the metal phase. We confirm this behavior by a numerical method and an analytic calculation. As a byproduct, we show the existence of a `quantum phase transition' supported by the Breitenlohner-Freedman bound argument.
\\ [15mm]
Keywords: Gauge/gravity duality, Holographic conductivity, Metal-insulator transition 

\vspace{5mm}
\end{titlepage}

\section{Introduction}
The metal-insulator transition is one of the oldest but not yet fully understood phenomena in condensed matter physics\cite{mott1968metal,imada1998metal,kravchenko2003metal,mott2004metal}. In conventional metals, electrons are elastically scattered by the atoms of the material but can move around quite freely. The electric conductivity is governed by the Drude model, and resistivity has $\rho \sim T^2 $ behavior in a two-dimensional system. As the electron-electron or electron-impurity interaction increases, the electric conductivity is no longer accounted for by the Drude model. One such example is the so-called strange metal whose electric resistivity increases linearly with temperature. This strange metal phase is thought to be closely related to the high $T_C$ superconducting phase.

On the other hand, freely moving electrons are rarely present in an insulator. As opposed to the phenomena in the metallic phase, the electric resistivity is very large at low temperatures and the resistivity decreases with increasing temperature. This temperature behavior is a main characteristic of insulators.

In condensed matter physics, there are several processes for the formation of the insulating phase. One simple example is the `band insulator'. In quantum mechanics, electrons in a periodic potential form a band structure. When the Fermi surface is in a gap between two bands, the electrons in the lower band (valence band) need some amount of energy to jump up to the upper band (conduction band). If the energy of electrons is not large enough to overcome this gap, the system remains insulating phase, a so-called `band insulator'. However, these band insulators cannot handle the correlation effect of electrons as they are explained by single-particle-picture-based quantum mechanics.

In this paper, we will focus on the insulating process induced by strong correlations, especially the electron-electron interactions and the electron-impurity interactions. The insulating process by the electron-electron interaction is known as the `Mott transition' \cite{mott1949basis}. In this case, each lattice site is occupied by an electron, and the electric current is generated by hopping electrons from one site to its nearest neighbor sites. When the electron-electron interaction becomes strong, that is, when the on-site electron potentials have a large barrier, the Coulomb repulsion is large, so that an electron in one site cannot hop to the next site. Therefore, all electrons are confined to their site. This is called the `Mott insulator'.

The other type of insulating mechanism is known as the `Anderson localization', which is governed by electron-impurity interaction \cite{abrahams1979scaling}. When the interaction between electrons and impurities increases, the electrons return to their original positions scattered by impurities. This effect enhances the wave function to confine the electron state to its original position. This type of insulator is called the `Anderson insulator'.

Due to the non-perturbative nature of the `Mott insulator' or `Anderson insulator', it is very hard to analyze the insulating process in a perturbative manner. The insulating process in the `Mott insulator' can be understood through the Hubbard model. However, if the system is complicated, it is very difficult to diagonalize the Hamiltonian matrix. The process of the `Anderson insulator' is much more complex because it depends on the details of the disorder\cite{kravchenko2003metal}.

For this reason, there are several studies describing the metal-insulator transition using gauge/gravity duality\cite{Mefford:2014gia, Baggioli:2014roa, Baggioli:2016oqk, Grozdanov:2015qia, Baggioli:2016oju, An:2020tkn,Ling:2016dck, Ling:2014saa,Donos:2012js}. The key physical quantity that determines a metallic or insulating phase is electric conductivity. In holography, we usually obtain electric conductivity by turning on the gauge field fluctuations around the background solution. Considering the infalling and regularity condition near the black brane horizon, we can get the boundary electric current and the electric conductivity\cite{Donos:2014cya,Amoretti:2014zha,Donos:2014yya}. This method has been applied to various extensions, {\it e.g.} \cite{Blake:2014yla,Kim:2014bza,Donos:2014oha,Davison:2014lua,Lucas:2014sba,Ge:2014aza,Kim:2015dna,Amoretti:2015gna,Kim:2015wba,Zhou:2015dha,Blake:2015epa,Donos:2015gia,Donos:2015bxe,Seo:2015pug,Kim:2016hzi,Seo:2016vks,Donos:2017oym,Kim:2019lxb,Kim:2020ozm, Jeong:2021wiu, Jeong:2018tua, Ahn:2019lrh}.

Holographic DC conductivity is divided into the charge conjugation symmetric part and the momentum dissipation part\cite{Blake:2014yla}. The former arises from the electron-hole pair creation, and the latter is contributed from the momentum dissipation by impurities. The momentum dissipation part of the electric conductivity is proportional to the square of the charge-carrier density and inversely proportional to the impurity density. This term can be suppressed when the impurity density is much greater than the charge-carrier density\cite{mott1949basis, mott2004metal , Kim:2014bza}. On the other hand, the charge conjugation symmetric part is proportional to the gauge coupling which is a coefficient of the Maxwell term $F^2$ in the action. This gauge coupling term is the lower bound of the DC electric conductivity\cite{Grozdanov:2015qia}. If the gauge coupling is constant, the DC conductivity cannot be lower than that value even at the zero temperature limit\cite{ Kim:2014bza, Baggioli:2014roa}.

One way to relax this bound is to set the gauge coupling as a function of another field or parameter so that the value can be small\cite{Mefford:2014gia, Baggioli:2016oqk}. In \cite{Mefford:2014gia}, the authors introduced a neutral scalar field in the bulk theory and the gauge coupling is a function of this scalar field. They also introduced Mexican hat-type scalar potential where the value of the scalar field makes the gauge coupling to be zero. Similar behavior can be observed using the axion field, which corresponds to the momentum relaxation in the boundary theory\cite{Baggioli:2016oqk}. In this paper, we are using a neutral scalar field inspired by \cite{Mefford:2014gia}. Instead of introducing a scalar potential, we investigate dynamical condensation in the bulk through interaction with the gauge field.

This paper is organized as follows. In section \ref{section:BG}, we introduce a new interaction term between the scalar and the gauge field, describing the order parameter and the charge-carrier density. We get the background geometry numerically and find a phase transition from RN-AdS black branes to hairy black branes. We also discuss an appearance of the `quantum phase transition'.\footnote{We do not consider hairy black branes at zero temperature, which is very difficult to deal with using our numerical method. However, we can find hairy solutions at  very close to zero temperature. Our argument is established in this sense.} In section \ref{section:DC}, we calculate DC electric conductivity by using a standard holographic method. We classify the phases based on the temperature dependence of the electric conductivity. In section \ref{section:diss}, we discuss our results and provide future directions.

\section{Background geometry and quantum phase transition}\label{section:BG}

In this section, we will discuss the background geometry of Einstein-Maxwell-dilaton with an axion field. We also introduce the dilation interaction term with the $U(1)$ gauge field.  We find two possible solutions to the equations of motion. Comparing the free energy of the solutions, we find phase transitions between the two solutions. We also discuss the `quantum phase transition' of the system.

We start from Einstein-Maxwell-dilaton action with an axion field,
\begin{align}\label{eq:Stot}
S_{tot} = S_0 + S_{int} + S_{bd}\,,
\end{align}
where
\begin{align}\label{eq:S0}
S_0 = \frac{1}{16\pi G} \int d^4 x \sqrt{-g} \left(R -2 \Lambda -\frac{1}{2} \sum_{\mathcal{I}}\left(\partial \chi^{\mathcal{I}} \right)^2 -\frac{1}{4}F^2 -\frac{1}{2}\left(\partial \phi\right)^2 -\frac{1}{2} m^2 \phi^2 \right),
\end{align}
and the interaction term is
\begin{align}\label{eq:Sint}
S_{int}= - \int \sqrt{-g} \frac{\gamma}{4} \phi^2 F^2\,.
\end{align}
Also, $S_{bd}$ consists of the Gibbons-Hawking term and counter terms for the holographic renormalization~\cite{Baggioli:2021xuv}. Here $\chi^{\mathcal{I}}$ is the axion field which gives a momentum relaxation effect on the boundary theory. In addition, the cosmological constant has been chosen as $\Lambda=-\frac{3}{L^2}$ for the asymptotic AdS geometry.
The equations of motion of the action (\ref{eq:Stot}) are
\begin{equation}\label{eq:eom}
\begin{split}
&R_{MN}-\frac{1}{2} g_{MN} {\cal L} - \frac{1}{2} \partial_M \phi \partial_N \phi -\frac{1}{2} \sum_{\mathcal{I}}\partial_M \chi^{\mathcal{I}} \partial_N \chi^{\mathcal{I}} -\frac{1}{2} \left( 1+\gamma \phi^2 \right) F_{MP}F^{P}_{M}=0\,, \cr
&\nabla^2 \phi -\left(m^2 +\frac{1}{2} \gamma F^2 \right) \phi =0\,, \cr
&\nabla_M \left(1+\gamma \phi^2\right) F^{MN} =0 \,,\cr
&\nabla^2 \chi^{\mathcal{I}} =0\,,
\end{split}
\end{equation}
where ${\cal L}$ is a Lagrangian density of $S_0$ and $S_{int}$. To solve the equations of motion, we take an ansatz as follows;
\begin{align}\label{ansatz}
&ds^2 = - U(r) e^{2(w(r)-w(\infty) )} dt^2 + \frac{r^2}{L^2} (dx^2+ dy^2) +\frac{dr^2}{U(r)}~,\nonumber\\
&A = A_t(r) dt~,~\chi^{\mathcal{I}} = \beta\,(x,y)~,~\phi=\varphi(r)\,.
\end{align}
With this ansatz, the equation of motion for the axion field $\chi^{\mathcal{I}}$ is automatically satisfied and $\beta$ turns out to be the dissipation of momentum via scattering with impurities. In addition, the Maxwell equation is integrable, so we define a conserved charge density as
\begin{align} \label{QQQ1}
\mathcal{Q} \equiv \frac{\sqrt{-g}}{L} (1 + \gamma \phi^2) F^{tr}\,.
\end{align}
We will regard this as a charge-carrier density below.

For the numerical computation, we rescale $r,~\mathcal{Q},~U,~\beta$ as

\begin{align}
\tilde{r} = \frac{r}{r_h },~\tilde{\mathcal{Q}} = \frac{L^4}{r_h^2}\mathcal{Q},~\tilde{U}(\tilde{r}) = \frac{L^2}{r_h^2} U(r),~\tilde{\beta} = \frac{L^2}{r_h} \beta\,,
\end{align}
such that all quantities are dimensionless. Then, the event horizon is located at $\tilde{r}=1$. From here, we omit the tilde not to clutter but note that all $r,~\mathcal{Q},~U,~\beta$ are scaled variables. Together with the Maxwell equation, the equations of motion for this hairy black brane can be written as follows:
\begin{align}\label{eom}
w'&-\frac{1}{4} r \varphi'^2 = 0\,,\cr
\varphi ''&+ \left(\frac{1}{r}-\frac{2 \beta^2 r^2+\frac{\mathcal{Q}^2}{\gamma \varphi ^2+1}-2 \left(\varphi ^2+6\right) r^4}{4 U\, r^3}\right)\varphi '+\frac{\gamma \mathcal{Q}^2 \varphi }{U\, r^4 \left(\gamma \varphi ^2+1\right)^2}+\frac{2 \varphi  }{U}=0\,,\cr
U'&+\frac{1}{4} U \, r \varphi'^2+\frac{\beta^2+2 U-\left(\varphi ^2+6\right) r^2}{2 r}+\frac{\mathcal{Q}^2}{4 r^3 \left(\gamma \varphi ^2+1\right)} =0\,,
\end{align}
where we set $m^2 =-2/L^2$.
For  $\varphi(r)=0$, we get RN-AdS black hole solution as;
\begin{equation}
\begin{split}
U(r) = \left(1-\frac{1}{r}\right)\left(1+r+r^2-\frac{\mathcal{Q}}{4r}-\frac{\beta^2}{2}\right)\,, \qquad w(r)=0\,.
\end{split}
\end{equation}
The entropy density and the temperature are given by
\begin{align}\label{eq:ST}
\begin{split}
s &= \frac{r_h^2}{4 G L^2} \,,\\
T &=  \frac{r_h}{4\pi L^2} e^{w(1)-w(\infty)} U'(1)~\,,
\end{split}
\end{align}
where $U'(1)$ can be written by using an equation of motion as follows:
\begin{align}\label{eq:temp}
U'(1) =  \frac{6+\varphi(1) ^2-\beta^2}{2}-\frac{{\cal Q}^2}{4  \left(1+\gamma \varphi (1)^2\right)}~.
\end{align}
Imposing the horizon regularity, $\varphi'(1)$ is required to be
\begin{align}
{\varphi}'(1)=\frac{4 \varphi(1)\left(\gamma {\cal Q}^2 +2(\gamma \varphi (1)^2+1)^2\right)}{(\gamma \varphi (1)^2 +1)({\cal Q}^2 +2 (\beta^2 -\varphi (1)^2 -6)(\gamma \varphi (1)^2 +1))}\,.
\end{align}
Hence, the solution of the equations of motion (\ref{eom}) can be parameterized by $(\gamma,~\beta,~{\cal Q},~\varphi(1))$\footnote{We will denote the horizon value of scalar field $\varphi(1)$ by $\varphi_h$. }. The asymptotic behavior of the scalar field near the boundary is
\begin{align}
\varphi(r) |_{r\rightarrow \infty} \sim \frac{J_{\varphi}}{r} +\frac{<{\cal O}_{\varphi}>}{r^2} +\cdots,
\end{align}
where the coefficient of the leading term denotes the source of the boundary operator ${\cal O}_{\varphi}$. In this work, we focus on the physics driven by the external electric field and dissipation only. Accordingly, we will take boundary condition $J_{\varphi}=0$.

In this work, we set $\gamma = -0.2$, which is a negative value for most calculations. As shown in the bulk action, the coefficient of $F^2$ term becomes $-\frac{1}{4}(1+ \gamma \phi^2)$, which should be negative for satisfying the null energy condition. Therefore, we have to check that the coefficient of the gauge field kinetic term $(1+  \gamma \phi^2)$ is positive for the given parameter range. If the coefficient becomes negative, then the gauge field fluctuation becomes ghost which leads to instability of the background geometry. In this paper, we focus on the ghost-free region of the gauge field fluctuation and leave comments in the discussion session.

\begin{figure}[ht!]
\begin{center}
\subfigure[$\beta=2.3$]
   { \includegraphics[width=6cm]{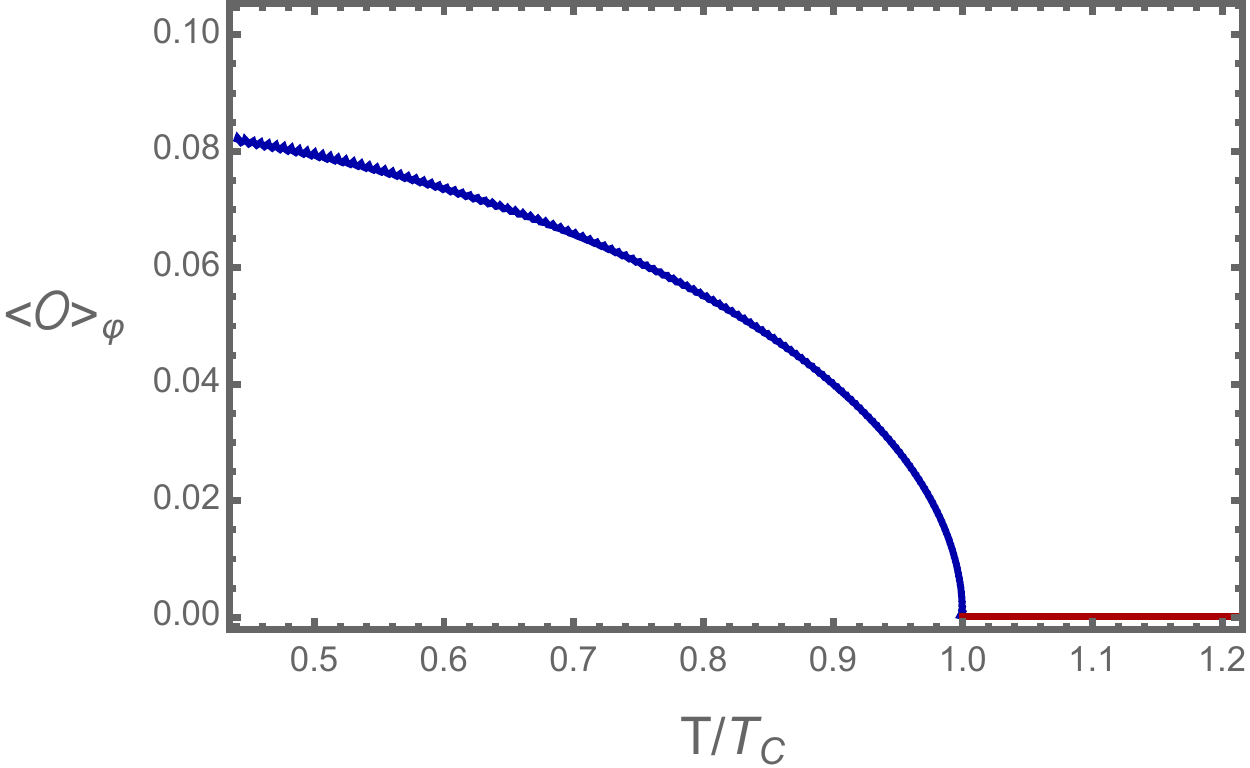}  }
\hskip.5cm   
\subfigure[${\cal Q}=0.5$]
   { \includegraphics[width=6cm]{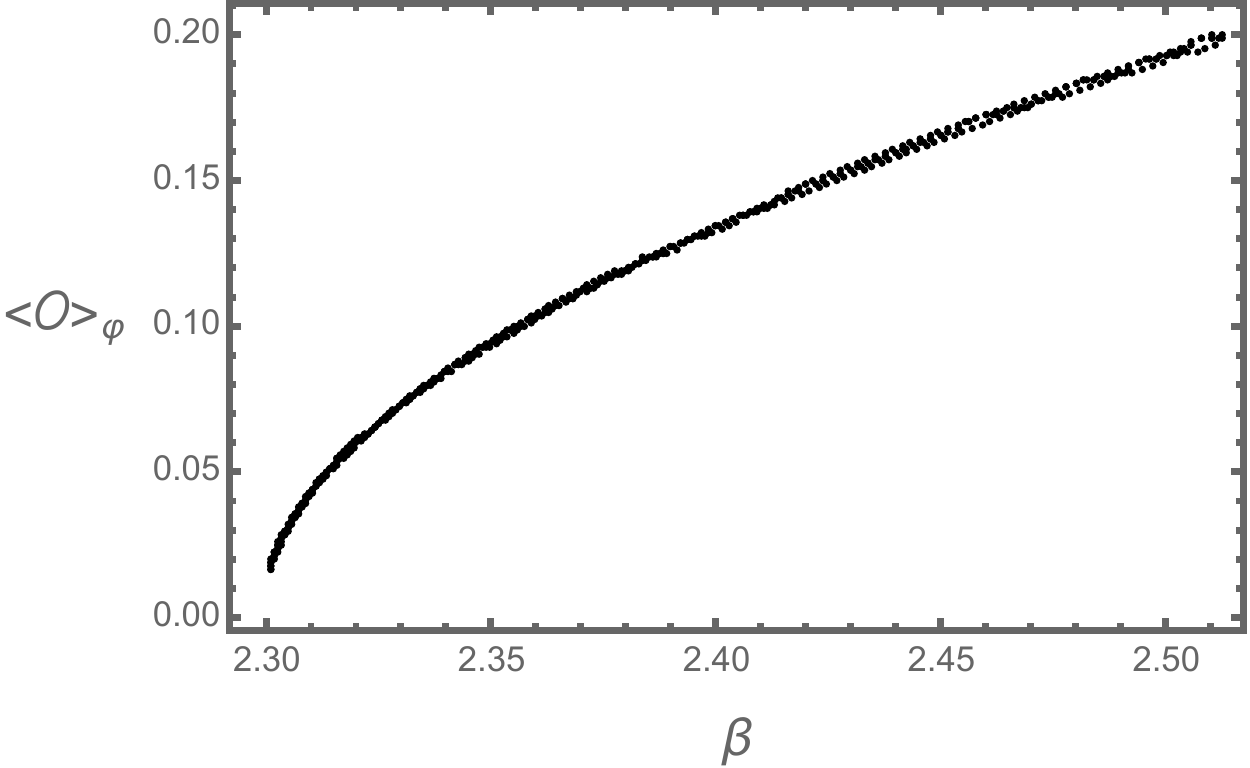}  }
    
\caption{(a) Temperature dependence of the scalar condensation. (b) Momentum relaxation parameter dependence of the scalar field condensation. In both cases, we set $\gamma =-0.2$. In this calculation, we use the canonical ensemble and take $r_h=1$ simply due to the convenience of the calculation.}
\label{fig:TO}
\end{center}
\end{figure}

Similar to the holographic superconductor model, we find the appearance of scalar condensation. At high temperatures, the background solution is the usual 4-dimensional AdS Reissner-Nordstrom (RN) black brane without scalar field(red line in Figure \ref{fig:TO} (a)). Since we checked that a black brane with scalar hair has smaller free energy than the RN-AdS black brane as the temperature decreases, such a scalar-hairy black brane is preferred at low temperatures. Hence, the hairy black brane geometry is adopted as a physical solution shown as a blue line in  Figure \ref{fig:TO} (a).

One interesting phenomenon of the model is the effect of momentum relaxation on scalar condensation. Figure \ref{fig:TO} (b) shows the momentum relaxation parameter $\beta$ dependence of the scalar condensation for a given charge density. As shown in the figure, as the momentum relaxation parameter increases, the value of the scalar condensation also increases. Therefore one can say that the momentum relaxation enhances the scalar condensation. In other words, the order parameter can be enhanced by impurities.

The enhancement of the scalar condensation by impurities appears to be non-trivial in this model. From the bulk action point of view, there is no direct interaction between the axion field and the real scalar field, which governs the order parameter. However, one can find that there is complicated mixing between all fields in the equations of motion level (\ref{eq:eom}). In particular, the axion field and the real scalar field appear together in the Einstein equation. See the first line in (\ref{eq:eom}). Therefore, we can expect two fields to interact through gravity. In the holographic description, this means that the order parameter can interact with impurities via charge-carrier exchange.

This phenomenon can be understood by (\ref{eq:ST}) and (\ref{eq:temp}). From the equations, temperature decreases as the momentum relaxation parameter $\beta$ increases. The value of scalar condensation increases as the temperature decreases. Therefore, the momentum relaxation parameter can enhance scalar condensation.  We observed similar phenomena in the different model \cite{Kim:2020ozm} in which the momentum relaxation parameter similarly lowers the temperature.

\begin{figure}[ht!]
\begin{center}
\subfigure[${\cal Q}=0$]
   { \includegraphics[width=6cm]{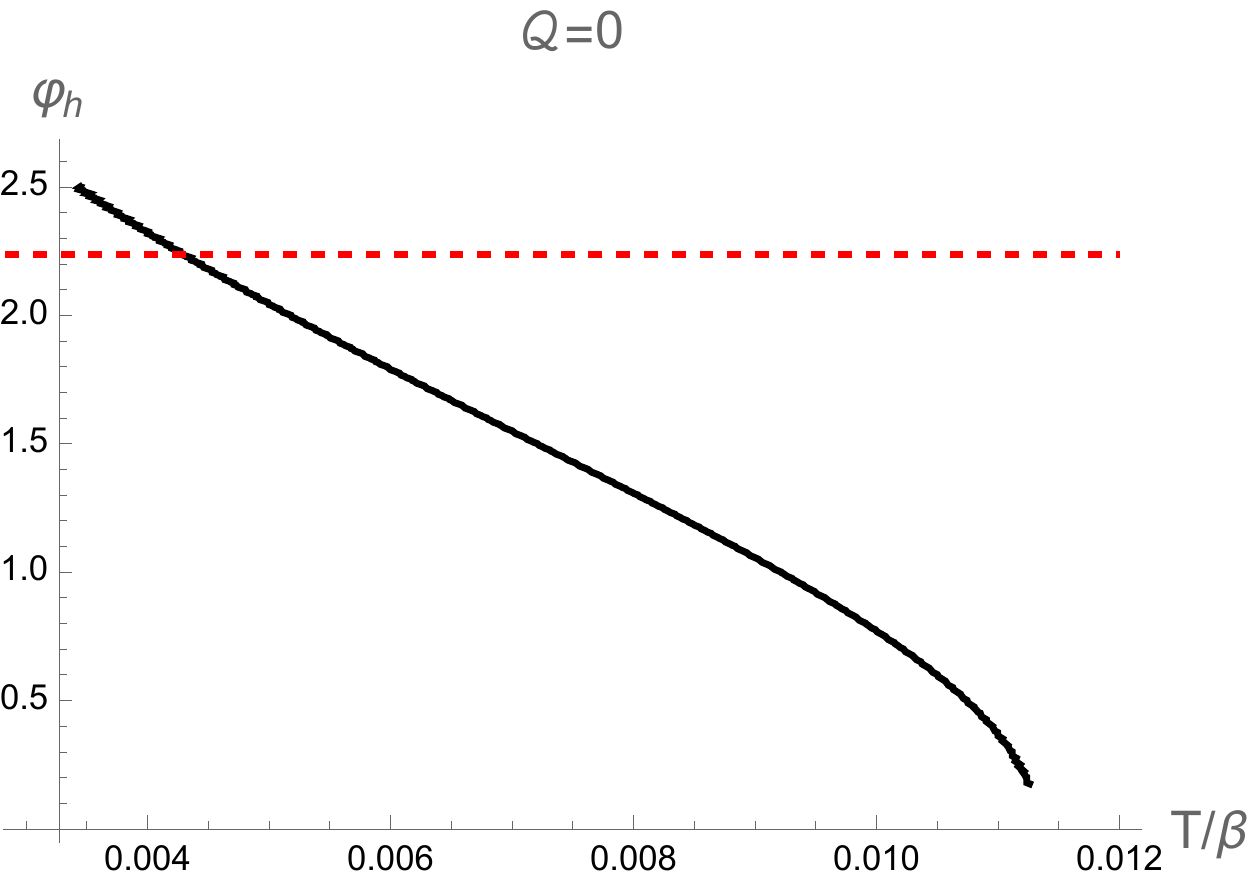}  }
\hskip.5cm   
\subfigure[${\cal Q}=0.01$]
   { \includegraphics[width=6cm]{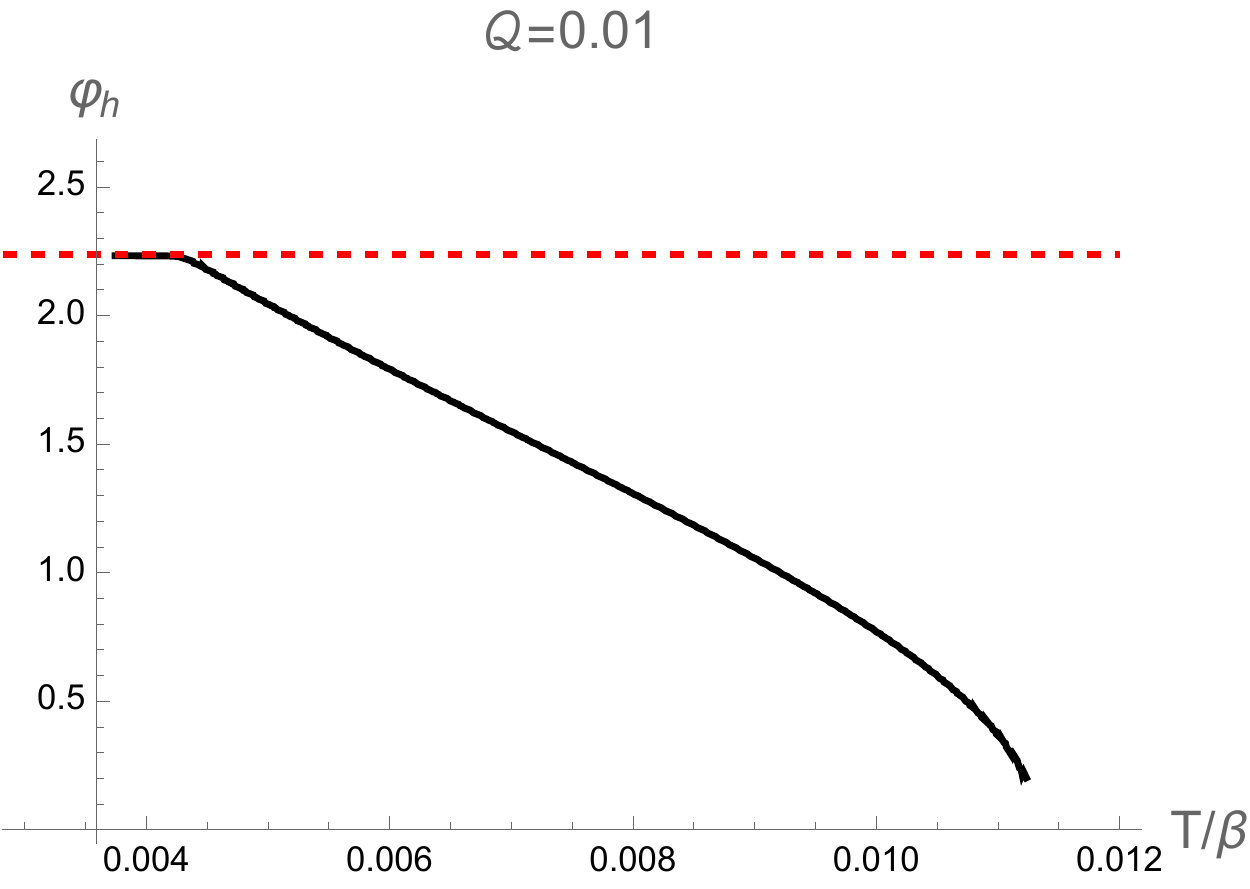}  }
    
\caption{Temperature dependence of the horizon value of the scalar field with the source free condition for (a) ${\cal Q}=0$ (b) ${\cal Q}=0.01$. The red dashed line indicates the sign change of the kinetic term of the gauge field.}
\label{fig:kineF}
\end{center}
\end{figure}

The appearance of the hairy black hole solution at low temperature is observed in the most range of charge density except ${\cal Q}=0$. Figure \ref{fig:kineF} (a) shows the temperature dependence of the horizon value of the scalar field, which satisfies the source-free condition. The red dashed line denotes $-1/\sqrt{\gamma}$. In the figure, the kinetic term of the gauge field changes sign at a certain temperature.  This phenomenon does not happen at finite charge density. See Figure \ref{fig:kineF} (b). As temperature decreases, the horizon value of the scalar field for the source-free condition increases. From the boundary theory point of view, it corresponds to the increasing expectation value of the scalar operator. However, the horizon value of the scalar field approaches $-1/\sqrt{\gamma}$. It does not increase and remains a constant value  as temperature decreases. It can be understood as follows. For the finite charge density case, if the horizon value of the gauge field is greater than $-1/\sqrt{\gamma}$, then the gauge field becomes tachyonic near horizon because of the wrong sign of the kinetic term of the gauge field. Therefore, the solution is not allowed from the equation of motion. On the other hand, zero charge density is  obtained by vanishing $A_{t}'(r_h)$ from (\ref{QQQ1}). In this case, the gauge field does not appear in the equations motion, and any value of $\varphi_h$ can be possible. To see the stability of the system, we have to take into account gauge field fluctuation, which will be discussed in the next section.

\begin{figure}[ht!]
\begin{center}
   { \includegraphics[width=7cm]{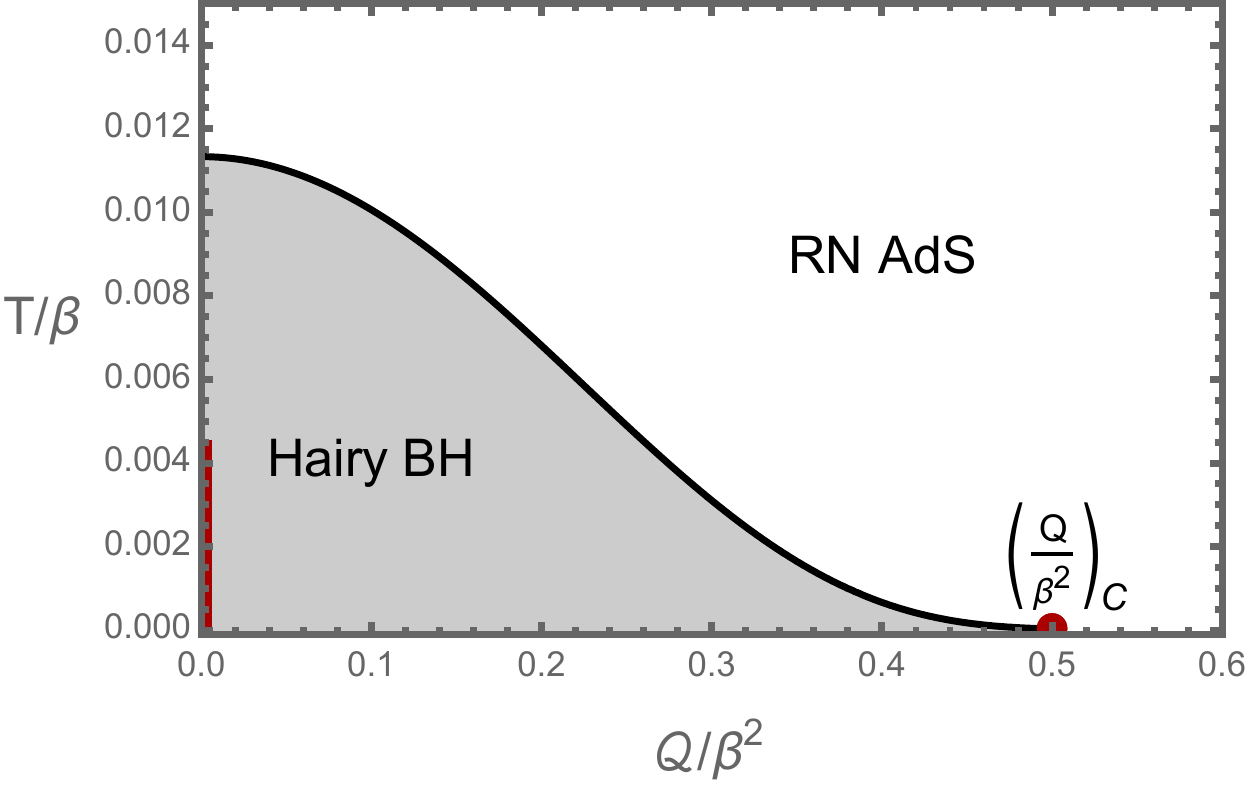}  }
       
\caption{The phase diagram of the system in a canonical ensemble with $\gamma =-0.2$.}
\label{fig:PT}
\end{center}
\end{figure}

The summarized phase diagram of the system in a canonical ensemble is shown in Figure \ref{fig:PT}. To make all parameters dimensionless, we scaled temperature and charge density by momentum relaxation parameter $\beta$.   In the figure, the system is divided into two regions. At high temperatures with large charge density, the normal RN-AdS black brane geometry without the scalar field is the only solution. On the other hand, at low temperatures with a small charge density, the hairy black brane solution with scalar hair becomes a preferable one. As we discussed earlier, $\beta$ (dissipation of momentum via scattering with impurities) enhances scalar condensation. In addition, the transition temperature between RN-AdS black brane and hairy black brane solution increases as the momentum relaxation parameter $\beta$ increases. In Figure \ref{fig:PT}, this behavior can be seen by decreasing $\mathcal{Q}/\beta^2$. The red line in ${\cal Q}/\beta^2 =0$ denotes the region where $(1+\gamma \varphi_h^2)$ becomes negative. We will discuss this region in the next section.

One interesting thing in the figure is that if the charge-carrier density is large enough to the impurity density(or fast dissipation region), the hairy black brane solution cannot exist at all temperatures. Therefore, the phase boundary is closed at zero temperature with finite charge density $ \left( \frac{\mathcal{Q}}{\beta^2} \right)_C$. This implies there exists a `quantum phase transition' point in the phase diagram. This cannot be observed in the model by turning off the $\gamma$ parameter.

In the absence of $\gamma$ coupling, the hairy black brane solution is realized by violating the Breitenlohner-Freedman (BF)  bound in the IR region. In pure $AdS_{d+1}$ geometry with a radius $L$, the BF bound of the scalar field mass can be determined by the reality condition for the scaling dimension of the scalar operator as 
\begin{align}
m_{BF}^2  \ge - \frac{d^2}{4 L^2}\,.
\end{align}
In the extremal 4-dimensional RN-AdS black brane geometry, the asymptotic geometry is $AdS_4$ while the near horizon geometry becomes $AdS_2 \times \mathbb{R}^2$. If we set the scalar mass to $m^2=-2/L^2$, the system seems to be stable from the boundary theory point of view, but it violates BF bound in deep IR and causes a transition to the hairy black brane solution. This phenomenon only depends on the mass of the scalar field, and hence the hairy black brane solution always exists at very low temperatures.

In our model, the effective mass of the scalar field near horizon is changed by $\gamma$ interaction in (\ref{eq:Sint}) as follows
\begin{equation}\label{eq:meff}
\begin{split}
m^2_{\text{eff}} &= m^2  + \frac{1}{2} \gamma F^2 \cr
&=\left\{
\begin{array}{c}
 m^2   \\
 m^2 - \gamma \frac{{\cal Q}^2 L^6}{r_h^4}   \\
\end{array}
\right.
\begin{array}{c}
 (r\to\infty) \\
(r\to r_h) \\
\end{array}\,,
\end{split}
\end{equation}
where we use the extremal RN black brane solution as the background geometry.
The effective mass is the same as the original scalar mass at the boundary, so it describes the same boundary system. However, the effective mass changes due to the charge density in the near-horizon region. In the model, we set $\gamma$ to be negative then the effective mass increases as the charge density increases. It implies that the instability near horizon due to the violation of the BF bound can be cured by increasing charge density. Therefore, the RN black brane solution can be stable from a certain large value of the charge density at zero temperature, and hence the `quantum critical point' can appear. When the absolute value of $\gamma$ is large, the instability is cured by small charge density ${\cal Q}$ from (\ref{eq:meff}). Therefore, the hairy black hole region will shrink to increase the absolute value of $\gamma$. The $\gamma$ dependence of the phase diagram is shown in Appedix \ref{AA}. The detailed BF bound analysis on the extremal RN black brane is discussed in Appendix \ref{QCP}.

\section{DC conductivity and metal-insulator transition}\label{section:DC}
In this section, we calculate the DC electric conductivity in the dual field theory using a holographic method. From the analysis of the DC conductivity with varying temperatures, we will discuss the electric properties of each phase.

\subsection{Holographic DC conductivity}
We can obtain holographic conductivities by solving fluctuation equations. To do this, we turn on the small fluctuation of metric, gauge field, and axion field around the background solution:
\begin{equation}
\begin{split}
ds^2_{(1)} &= 2 \lambda  \left[ \delta g_{tx} (r) \, dt\, dx + \delta g_{ty} (r)\, dt\,dy +r^2 \delta h_{rx}(r) \, dr\,dx +r^2 \delta h_{ry}(r)\,dr\,dy \right] \,,\cr
A_{(1)} &= \lambda \left[ \left\{ - E_{x} t + \delta A_{x} (r) \right\} dx +\left\{ - E_{y} t + \delta A_{y} (r) \right\} dy \right] \,,\cr
\chi^{\cal I}_{(1)} &= \lambda \left( \delta \chi_{x} (r), ~\delta \chi_{y} (r) \right)\,,
\end{split}
\end{equation}
where $\lambda$ is a formal expansion parameter. From the vector property of the fluctuation, one can check that the fluctuation of the real scalar field $\varphi$ is completely decoupled at the linear level of $\lambda$. The linearized fluctuation equations are
\begin{equation}\label{eq:fluc}
\begin{split}
&\delta g_{ti}''(r) -\frac{r}{4} \varphi'(r)^2 \delta g_{ti}'(r) +\left( \frac{1}{2}\varphi'(r)^2 -\frac{\beta^2}{r^2 U(r)}-\frac{2}{r^2}\right) =0 \,,\cr
&\delta A_{i}''(r) - {\cal F}(r) \delta A_{i}'(r) +\frac{{\cal Q}e^{W(r)/2}}{r^3 U(r)(1+\gamma \varphi (r)^2)}\left(r \delta g_{ti}'(r)-2 \delta g_{ti}(r) \right) =0 \,,\cr
&\delta \chi''(r) + \left(\frac{2}{r}-\frac{2 \gamma \varphi(r)\varphi'(r)}{1+\gamma \varphi(r)^2} -{\cal F}(r)\right)\delta\chi'(r) -\frac{\beta}{r^2 U(r)}\delta g_{ti}'(r) -\frac{\beta \varphi'(r)^2}{4 rU(r)} \delta g_{ti}(r) =0\,,
\end{split}
\end{equation}
where
\begin{equation}
\begin{split}
&{\cal F}(r)\\
& = \frac{4r^2 U(1+\gamma \varphi^2 -3 \gamma r \varphi\varphi')-2 \gamma r^4 \varphi^4 -2r^2(r^2+6\gamma r^2 -\gamma  \beta^2)\varphi^2+{\cal Q}^2 -12r^4 +2 r^2 \beta^2}{4 r^3 U(1+\gamma \varphi^2)}\,,
\end{split}
\end{equation}
and $i$ denotes $x$ and $y$ directions.

Together with the regularity and the ingoing condition at the horizon, the behavior of the fluctuation of each field near horizon can be expressed as
\begin{align}
&\delta g_{ti}(r) \sim \delta g_{ti}^0 + \cdots,~~~~~~~\delta h_{ti}(r) \sim \frac{1}{r^2 U(r)} \delta g_{ti}^0 +\cdots ,\cr
&\delta A_{i}(r) \sim -\frac{E_{i}}{4\pi T} \log (r-r_h) + \cdots,~~~~~~\delta \chi^{{\cal I}} \sim \delta \chi_{0}^{{\cal I}}+\cdots.
\end{align}
On the other hand, the gauge field fluctuation near the boundary can be expressed as
\begin{align}
\delta A_i (r) \sim - E_i \, t + \frac{J^i}{r} + \cdots,
\end{align}
where the electric current $J_i$ is a response to the external source $E_i$.

Now, we  define conserved current ${\cal J}^i$ as
\begin{align}
{\cal J}^i&= \sqrt{-g} (1+\gamma \varphi^2) F^{ir} \cr
&= -U(r) (1+ \gamma \varphi(r)^2) \delta A_i'(r) - a_t'(r) \delta g_{ti}(r)\,.
\end{align}
This current ${\cal J}^i$ is independent of radial direction according to the equation of motion and it becomes electric current $J^i$ at the boundary. Therefore, we can obtain DC conductivity  in terms of the horizon data as
\begin{align}\label{eq:DC}
\sigma_{DC} = \left(1+ \gamma \varphi_h^2 \right) + \frac{e^{W(\infty)}{\cal Q}^2}{r_h^2 \beta^2}\,,
\end{align}
where $\varphi_h$ is the horizon value of the scalar field. The first two terms in (\ref{eq:DC}) are seemingly independent of charge density. These terms are understood as a consequence of the electron-hole pair creation by charge conjugation symmetry($\sigma_{ccs}$). The last term is proportional to the square of the charge density and inverse of the impurity density which refers to current dissipation by impurity or lattice($\sigma_{diss}$)\cite{Blake:2014yla}. Then, DC conductivity can be written as
\begin{align}\label{eq:DC3}
\sigma_{DC} = \sigma_{ccs} + \sigma_{diss}\,.
\end{align}

The dissipation part of the electric conductivity is the same as in usual holographic models. But the charge conjugation symmetry part $\sigma_{ccs}$ contains the horizon value of the scalar field which gives finite condensation of the scalar field. With a negative value of $\gamma$,  $\sigma_{ccs}$ term can be suppressed by scalar condensation. We will first discuss this suppression of the charge conjugation symmetry part of the conductivity, and then we will consider the full DC conductivity.

\subsection{DC conductivity with zero charge-carrier density}

In this section, we discuss DC conductivity without charge-carrier density (${\cal Q} =0$). The background solution now is nothing but the Schwarzschild black brane with momentum relaxation. However, the interaction term between the $U(1)$ gauge field and the real scalar $\varphi$ can exist at the fluctuation level and affects DC conductivity.

In the absence of charge-carrier density, there is no dissipation term in (\ref{eq:DC}) and hence only electron-hole pair creation contributes to the DC conductivity,
\begin{align}
\Big\{
\begin{array}{ll}
\sigma_{DC}\Big|_{{\cal Q} =0} &= 1+ \gamma \varphi_h^2  ~~{\rm for~hairy~BH}\,,\\
\sigma_{DC}\Big|_{{\cal Q}=0} &= 1~~~~~~~~~~~~{\rm for~RN~AdS~BH}\,.\\
\end{array}
\end{align}
Here, one can easily notice that the DC conductivity solution can be smaller than $1$ with a negative value of $\gamma$ in the presence of scalar condensation(for the hairy black brane). Moreover, the DC conductivity can be negative if $\gamma \varphi_h^2 < -1$. However, in this case, the kinetic term of $U(1)$ gauge field in the action changes the sign which implies that the gauge field fluctuation becomes a ghost near the black brane horizon. This ghost fluctuation of the gauge field causes instability near horizon and we speculate it leads to a geometrical phase transition, which could have an important physical implication for this model. In this work, we will not discuss this geometrical transition and postpone it to future work.

\begin{figure}[ht!]
\begin{center}
\subfigure[]
   { \includegraphics[width=7cm]{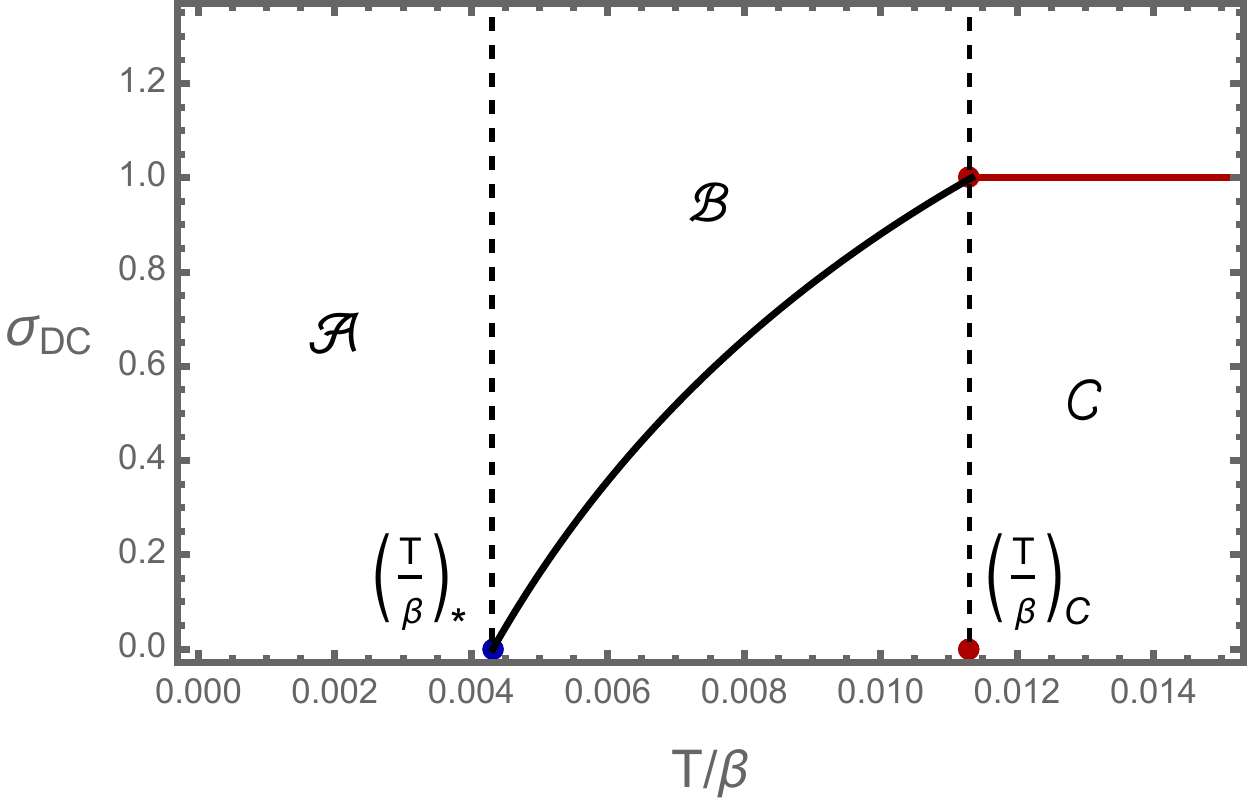}  }
   \hskip0.5cm
 \subfigure[]
   { \includegraphics[width=7.4cm]{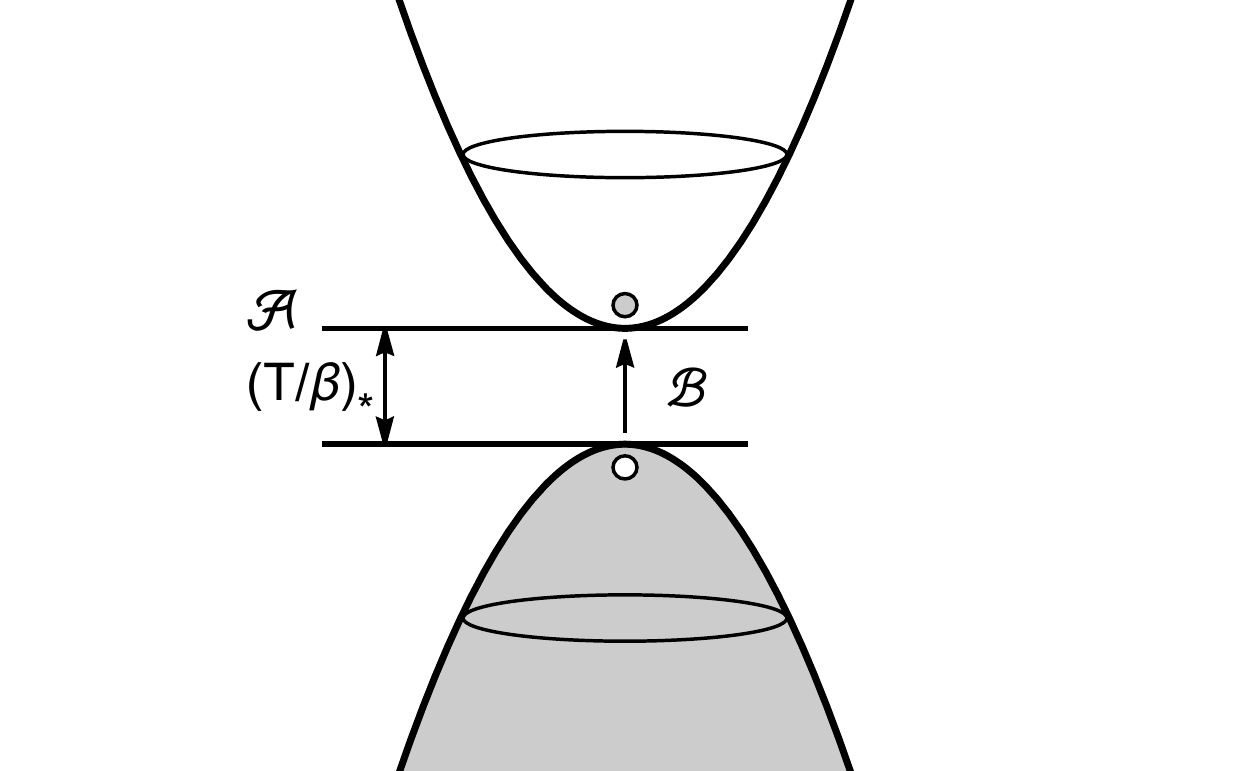}  }

\caption{(a) DC conductivity with vanishing charge-carrier density and $\gamma = -0.2$. (b) Interpretation of region ${\cal A}$ and ${\cal B}$ in the boundary system.}
\label{fig:DC0}
\end{center}
\end{figure}

The temperature dependence of DC conductivity without charge-carrier density is drawn in Figure \ref{fig:DC0}(a). The resultant DC conductivity can be divided into three regions. At high temperatures(region ${\cal C}$, $T/\beta>(T/\beta)_C$), the DC conductivity shows the typical behavior of the conductivity from the Schwarzschild black brane. This region continued to the metallic phase when charge-carrier density is added which will be discussed in the next section. In the intermediate temperature region (region ${\cal B}$, $(T/\beta)_* < T/\beta < (T/\beta)_C)$), the DC conductivity decreases as the temperature is lowering$(d \sigma_{DC}/d T >0)$ which indicates that the region ${\cal B}$ is the insulating phase. At low-temperature region (region ${\cal A}$, $T/\beta<(T/\beta)_{*}$), DC conductivity becomes negative which leads to the instability of the background as we discussed earlier.

The presence of $(T/\beta)_{*}$ indicates that there is a gap in the electron state. The schematic picture for regions ${\cal A}$ and ${\cal B}$ is drawn in Figure \ref{fig:DC0}(b). Since there is no charge-carrier density, we assume that the lower band is filled with electrons and the upper band is empty. If the temperature is lower than the gap energy, the corresponding DC conductivity vanishes because no electron-hole pairs can be created by thermal fluctuations. On the other hand, when the temperature reaches the gap energy and the thermal fluctuations overcome the gap, electron-hole pairs are created. The electrons move in the direction of the electric field and holes move in the opposite direction. Due to the opposite charges of electron and hole, a current can be generated in the same direction of the electric field, thus a nonvanishing DC conductivity can be obtained in the region $\mathcal{B}$.

When $\gamma$ coupling is turned on, a gap is created. As mentioned before, this can be seen from the fact that the DC conductivity becomes zero at a certain energy scale. This gap scale $(T/\beta)_*$ also increases as $\gamma$ is increasing. In the region $\mathcal{B}$, DC conductivity is monotonically increasing to the temperature. This implies the resistivity is decreasing to temperature, $\partial \rho /\partial T <0$, which is a typical characteristic of the insulator. Therefore, we expect that the dual system of the hairy black brane solution with $\gamma$ interaction is in the insulating phase.  The DC conductivity with different $\gamma$ is discussed in Appendix \ref{AA}.

\subsection{DC conductivity at finite charge density}

In the presence of the charge-carrier density, DC conductivity is consist of the charge conjugation symmetry part and dissipation part as in (\ref{eq:DC}) and (\ref{eq:DC3}),

\begin{align}\label{eq:DC2}
\Big\{
\begin{array}{ll}
\sigma_{DC} &= \left(1+ \gamma \varphi_h^2 \right) + \frac{e^{W(\infty)}{\cal Q}^2}{r_h^2 \beta^2} ~~~{\rm for ~ hairy~BH}\,,\\
\sigma_{DC} &= 1 + \frac{{\cal Q}^2}{r_h^2 \beta^2}~~~~~~~~~~~~~~~~~~~~{\rm for~RN~AdS~BH}\,.
\end{array}
\end{align}

When the temperature is low enough($ T/\beta < (T/\beta)_C$), the hairy black brane is a physical solution. Also, the corresponding DC conductivity becomes the first line of (\ref{eq:DC2}). On the other hand, RN-AdS black brane is a physical solution at high temperature and hence we get standard DC conductivity for an RN black brane as the second line of (\ref{eq:DC2}). In the case of the DC conductivity for a hairy black brane, we use a numerical solution with a source-free condition for a real scalar field.

\begin{figure}[ht!]
\begin{center}
   { \includegraphics[width=7cm]{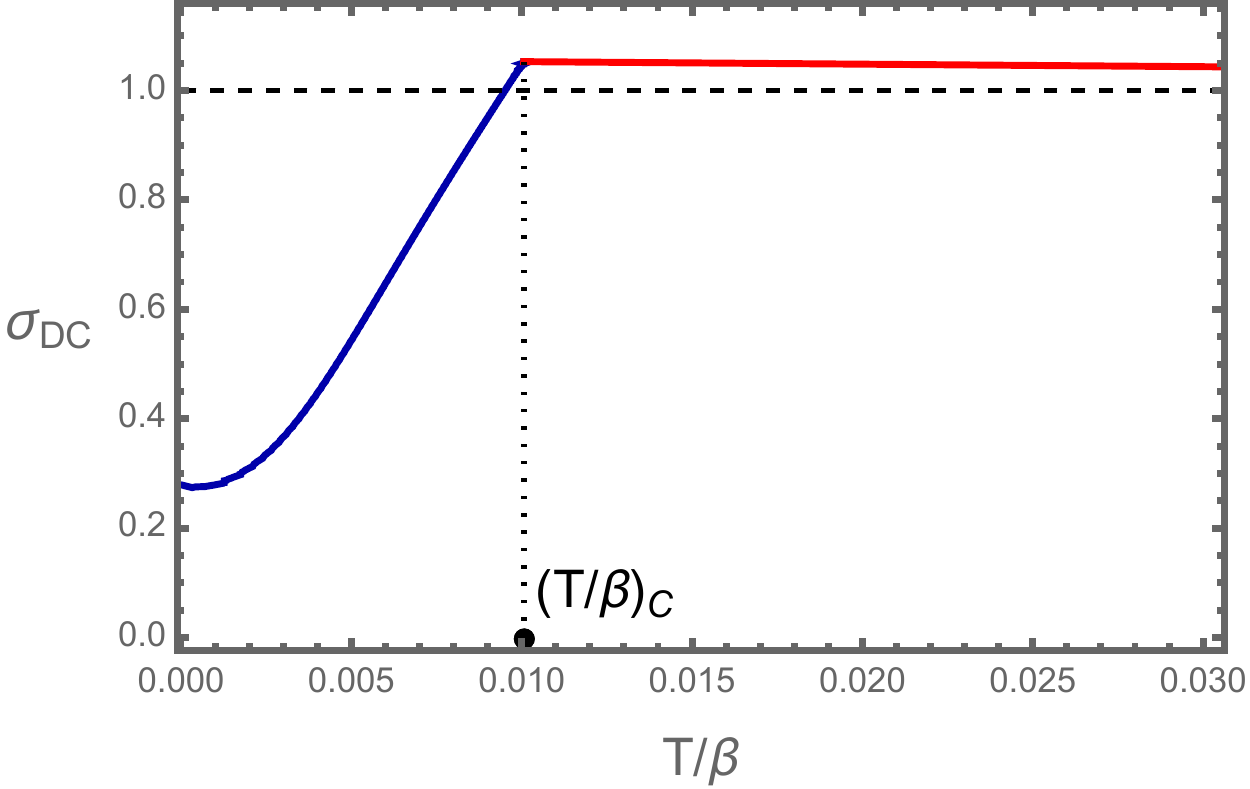}  }
   \hskip0.5cm

\caption{Temperature dependence of DC conductivity for  ${\cal Q}/\beta^2=0.1$ and $\gamma=-0.2$.}
\label{fig:DC2}
\end{center}
\end{figure}

 Figure \ref{fig:DC2} shows the temperature dependence of DC conductivity for given ${\cal Q}/\beta^2 =0.1 $ and $\gamma =-0.2$. The blue and red lines denote DC conductivities in hairy black brane solution and RN-AdS black brane solution, respectively. There are two features for the DC conductivity of hairy black branes. One is that the DC conductivity decreases as the temperature is lowered. This indicates that the dual system of the hairy black brane behaves like an insulator($\partial \rho/\partial T <0$) similar to the zero charge density case. The other one is that the DC conductivity goes to a finite value when the temperature approaches zero. It is due to the existence of charge-carrier density ${\cal Q}$ in the system. 

\begin{figure}[ht!]
\begin{center}
\subfigure[]
   { \includegraphics[width=7cm]{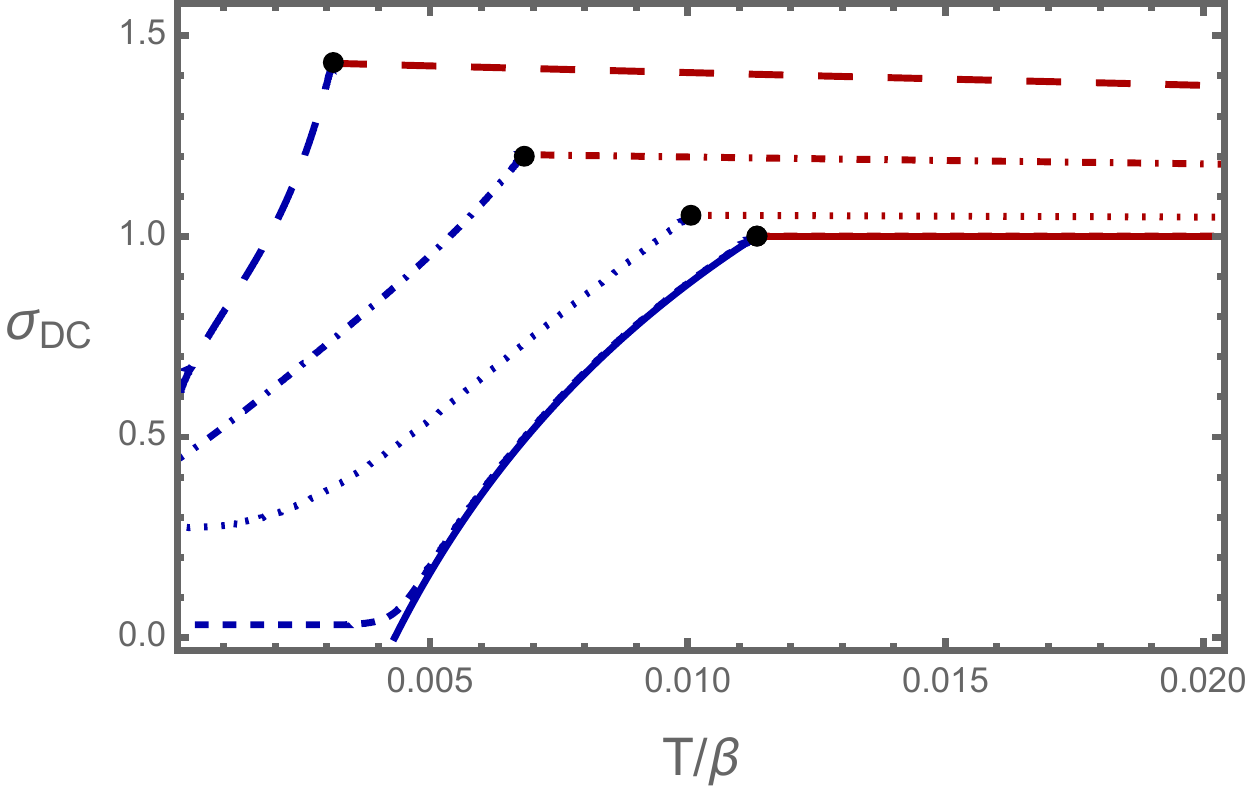}  }
   \hskip0.5cm
 \subfigure[]
   { \includegraphics[width=7cm]{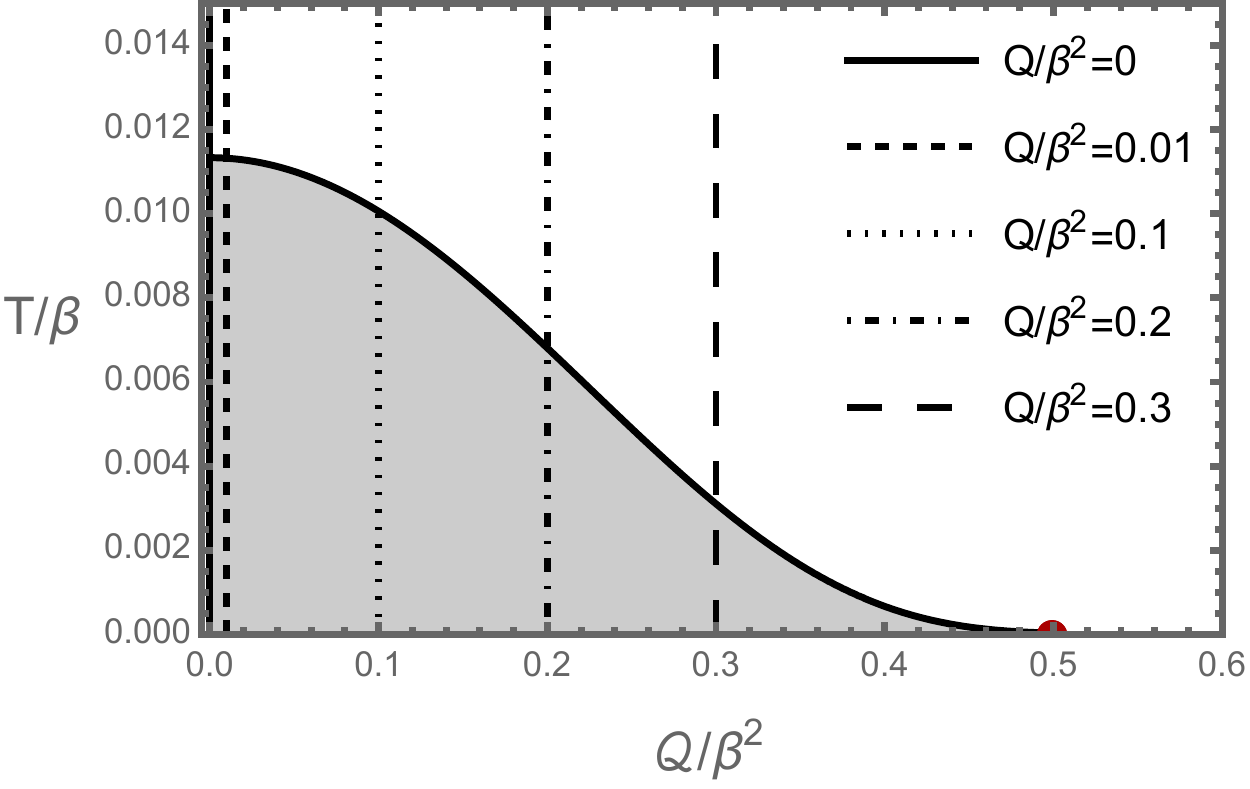}  }

\caption{(a) Temperature dependence of the DC conductivity along fixed charge density lines. (b) Fixed charge density lines correspond to (a).  Here, we set $m^2 =-2/L^2$ and $\gamma = -0.2$. }
\label{fig:DC3}
\end{center}
\end{figure}

After phase transition to RN-AdS black brane at $(T/\beta)_C$, DC conductivity behaves differently to the hairy black brane case. DC conductivity decreases as the temperature increases. It is not visible clearly in Figure \ref{fig:DC2} and \ref{fig:DC3}. However, if we fix the ratio $\mathcal{Q}/\beta^2$ based on these figures, the varying DC conductivity of the RN black brane can be written as
\begin{align}
\frac{d\sigma_{\text{DC}}}{d\left( T/\beta \right)} =-\frac{32 \pi  r_h \mathcal{Q}^2}{\beta \left(12 r_h^4+2 \beta ^2 r_h^2+3 \mathcal{Q}^2\right)} \,,
\end{align} 
where we took $L=1$ for simplicity. This expression shows the negative slope of the DC conductivity with respect to the temperature. 

The overall temperature dependence of DC conductivity in both the insulating and metallic phases is shown in Figure \ref{fig:DC3} (a). In the figure, blue lines correspond to the DC conductivity for the insulating phase and red ones to the metallic phase. The charge density corresponding to each line is shown in Figure \ref{fig:DC3} (b). In the figure, DC conductivity monotonically increases in the insulating phase while it decreases in the metallic phase. 


\begin{figure}[ht!]
\begin{center}
   { \includegraphics[width=6.5cm]{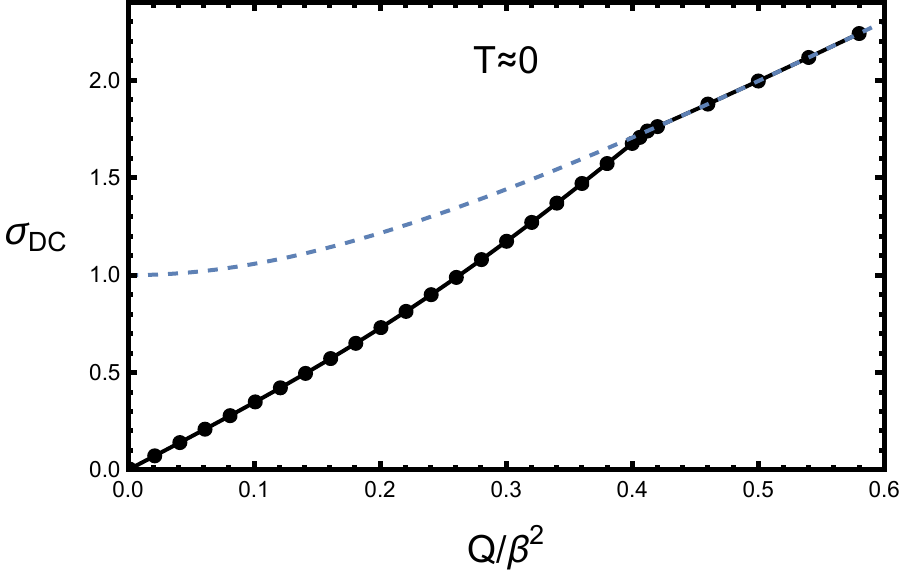}  }

\caption{The charge density dependence of DC conductivity at zero temperature for $\gamma=-0.2$. The dashed line indicates the DC conductivity in RN-AdS black brane solution.}
\label{fig:DC4}
\end{center}
\end{figure}

To see the physical properties of each phase, we calculate DC conductivity near zero temperature with $\gamma=-0.2$. See Figure \ref{fig:DC4}. In the figure, the dashed line denotes DC conductivity in the metallic phase showing quadratic behavior in charge density. The point where the conductivity line leaves the dashed line is the phase transition point to the insulating phase. In the insulating phase, the DC conductivity decreases faster than in the metallic phase as the impurity density becomes higher than the charge-carrier density. In the fast dissipation limit(${\cal Q}/\beta^2 \ll 1$), DC conductivity approaches zero for finite $\gamma$ interaction(The $\gamma$ dependence of the conductivity is shown in Appendix \ref{AA}). It implies that not only the dissipative part in the DC conductivity (\ref{eq:DC2})  but also the charge conjugation symmetry part of the DC conductivity are suppressed. Therefore, the low-temperature and fast dissipation regions can be interpreted as an insulating phase driven by the impurity, and this phase is speculated to be an Anderson insulator.

\begin{figure}[ht!]
\begin{center}
   { \includegraphics[width=7cm]{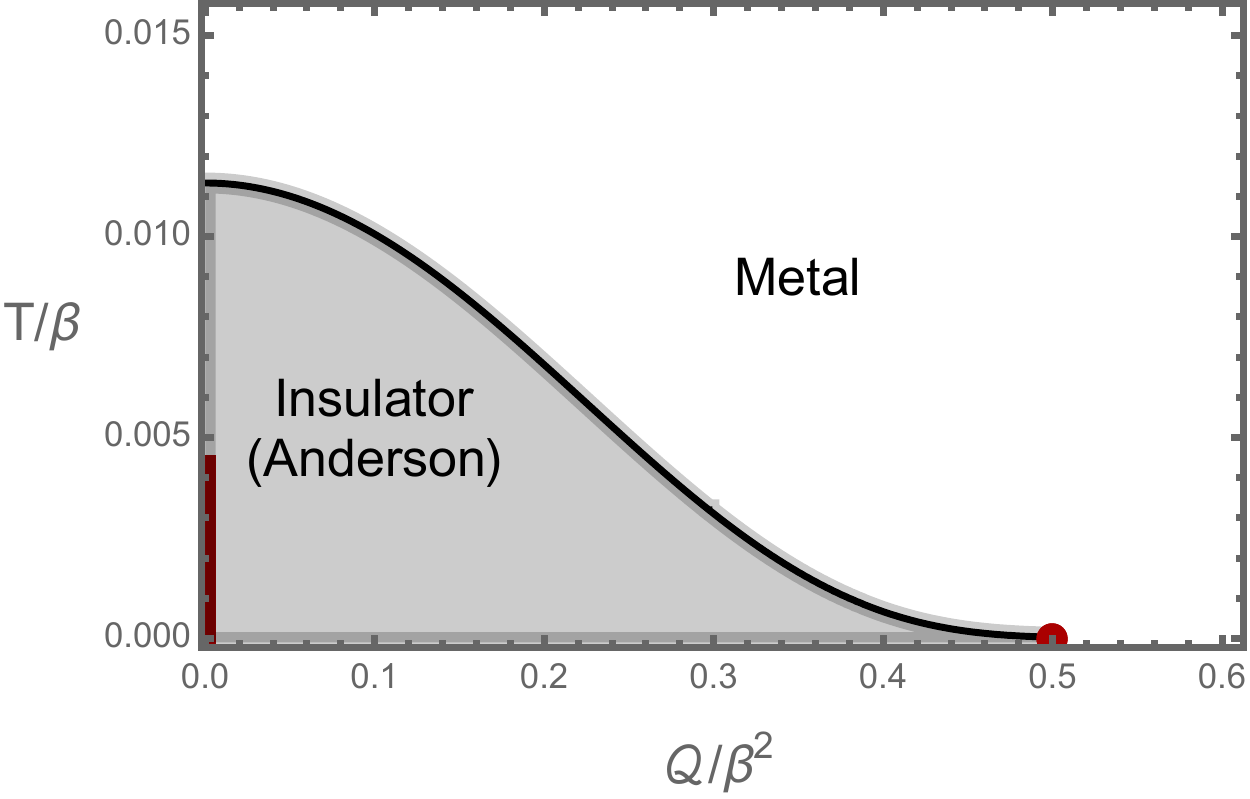}  }
   
\caption{The finally adopted phase diagram of the metal-insulator transition. }
\label{fig:PTF}
\end{center}
\end{figure}

Figure \ref{fig:PTF} is the finally adopted phase diagram of the system. In the regime of high temperature and high density(or low impurity density), the gravity system is RN-AdS black brane. This geometry corresponds to the metallic phase, whose resistivity increases with temperature.  On the other hand, at low temperatures and high impurity density(or low charge density), the geometry becomes a hairy black brane and the boundary system shows insulating behavior. In the insulating phase, both charge conjugation symmetry and dissipation parts of the electric conductivity are suppressed by impurity density which is very similar to the Anderson localization. We speculate this insulating phase corresponds to the Anderson insulating phase.  The red line along the temperature axis in Figure \ref{fig:PTF} denotes the region where our analysis is not applicable due to the unitarity violation of the gauge field fluctuation. We are expecting there should be another background phase transition to the non-black brane geometry. The resulting geometry might correspond to the Mott insulating phase. But we postpone this to future work.
\section{Discussion}\label{section:diss}

In this paper, we study a boundary system that undergoes the metal-insulator transition with an order parameter using the gauge/gravity duality setup. The corresponding gravity system is the Einstein-Maxwell-Axion theory with a neutral scalar field. This scalar field plays the role of the order parameter of the boundary system. We introduce a new interaction term between the scalar field and $U(1)$ gauge field. We set interaction strength $\gamma$ to be a negative number such that the effective charge of the gauge field is reduced. With this interaction term, we found a geometric phase transition between RN-AdS black brane and the hairy black brane solution. One remarkable result is that there exists a phase transition at zero temperature. This transition comes from the recovery of the BF bound near the extremal black brane horizon with a charge density. From the boundary theory point of view, the scaling dimension of the order parameter does not change. In other words, the UV theory does not change as varying charge-carrier density. In contrast, such a scaling dimension changes with charge-carrier density in the IR region. It leads to the `quantum phase transition' in the boundary theory.

We also calculate DC conductivity using the standard holographic method. We found that the electric conductivity in the hairy black brane phase increases with temperature. This behavior is known as a typical characteristic of insulators. Moreover, the suppression of the charge conjugation symmetry part and the dissipation part of the electric conductivity are induced by order parameter $\left<{\cal O}\right>_{\varphi}$ and impurity density $\beta$.  The order parameter  $\left<{\cal O}\right>_{\varphi}$ is also enhanced by impurity density. Therefore, we speculate that this insulating phase corresponds to the Anderson insulator phase raised by the localization due to impurity or disorder.  We don't have a clear interpretation of the order parameter in dual field theory. However, this insulating phenomenon looks different from Mott insulator physics. We need more study for the identification of this order parameter.

The phase diagram has a small window where the electrical conductivity becomes negative. In this region, the $U(1)$ gauge field becomes tachyonic near horizon, so the obtained solutions are not physically meaningful. In this region,  we expect that there should be a geometric phase transition to non-black brane geometry. The red line in Figure \ref{fig:PTF} does not carry any charge-carrier density.  It could correspond to a horizonless geometry. Then, the corresponding phase can be identified with the Mott insulator phase. Our speculated identification for the insulating phase is shown in Table \ref{table1}.

\begin{table}[htp]
\begin{center}
{\def\arraystretch{1.5}
\begin{tabular}{|c||c|c|}
\hline 
 
 & Physical origin & Holographic realization \\
\hline \hline
Band insulator & Periodic structure & Fermionic spectral function \\
\hline
Anderson insulator & Impurity induced & Hairy BH with scalar condensation \\
\hline
Mott insulator & Interaction induced & Horizonless geometry? \\
\hline
\end{tabular}
}
\end{center}
\caption{Holographic realization of insulating phases.}
\label{table1}
\end{table}%

We also study scaling behavior of the temperature dependence of resistivity in metallic and insulating phase. See Appendix \ref{scaling}.

Now, let us mention possible future directions. We expected that there must be a gravity dual to the Mott insulator in a region with small charge densities. A candidate for this gravity dual is a horizonless geometry with the axion field. Therefore, finding this solution can help us to understand the Mott insulation in a holographic study. In addition,  one may try to analyze AC conductivities on the hairy black branes for more concrete evidence. Also, it is interesting to introduce a complex scalar rather than a real scalar. There is a `quantum phase transition' from the charge of the scalar field. See \cite{Denef:2009tp,Hartnoll:2009sz,Kim:2015dna}. This consideration can show the metal-superconductor phase transition with a charged order parameter. We hope to report such a study near future \cite{OnGoing}.

\begin{appendices}

\section{Different $\gamma$ case}\label{AA}

\begin{figure}[ht!]
\begin{center}
\subfigure[]
   { \includegraphics[width=7cm]{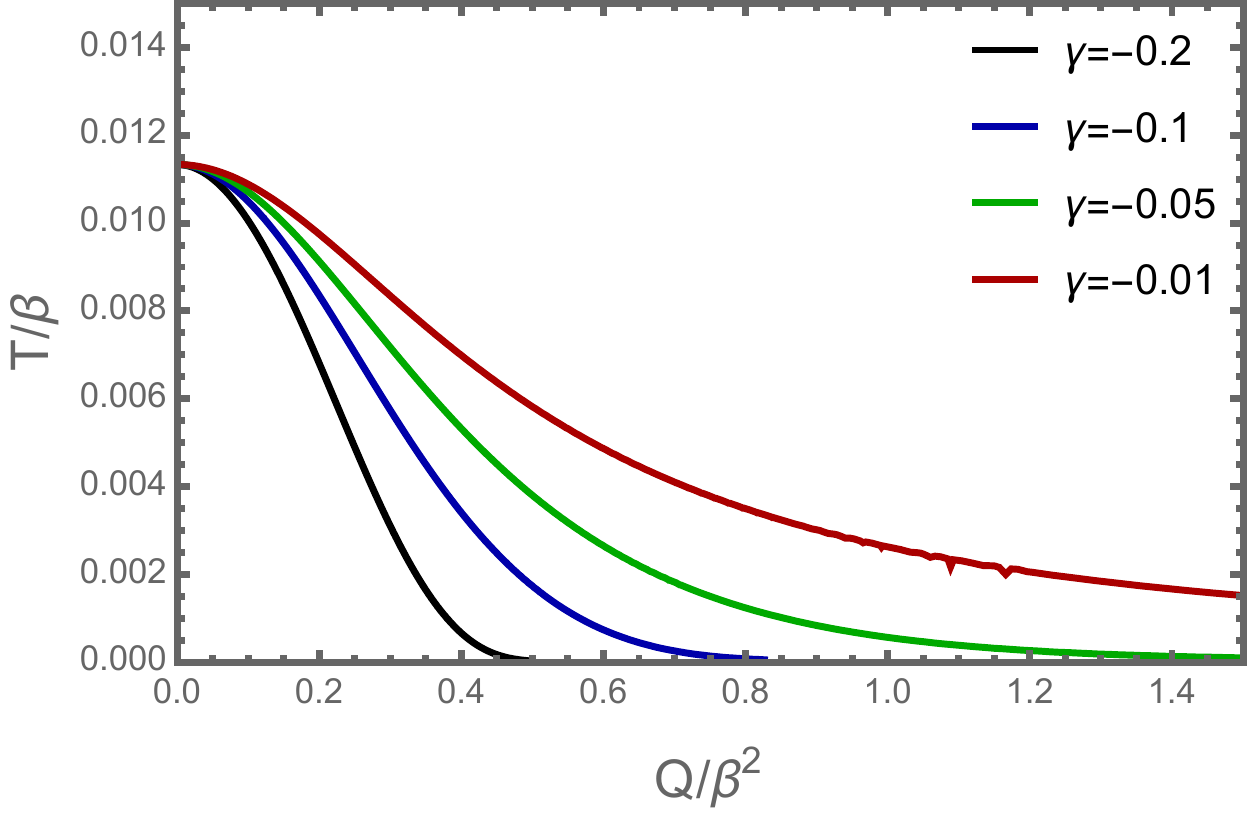}  }
   \hskip0.5cm
 \subfigure[]
   { \includegraphics[width=7cm]{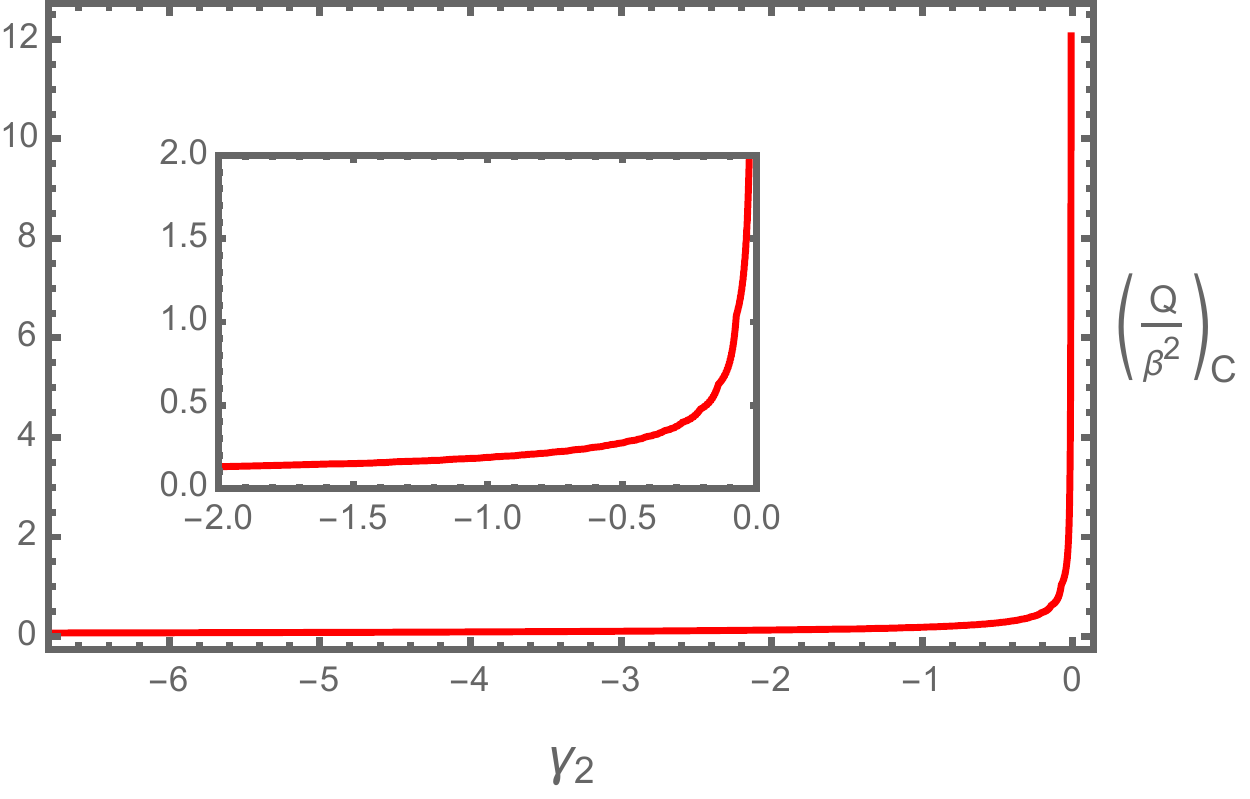}  }

\caption{(a) The phase diagram of the system with different $\gamma$. (b) $\gamma$ dependence of `quantum phase transition' point.}
\label{fig:g2PT}
\end{center}
\end{figure}

The effect of $\gamma$ on the phase diagram is shown in Figure \ref{fig:g2PT}(a). In this figure, each line denotes the phase boundary between the hairy black brane and RN-AdS black brane. As $\gamma$ goes to zero, the critical charge density of the `quantum phase transition' increases, and it seems to go to infinity when $\gamma =0$, see Figure \ref{fig:g2PT}(b).

\begin{figure}[ht!]
\begin{center}
\subfigure[]
   { \includegraphics[width=7cm]{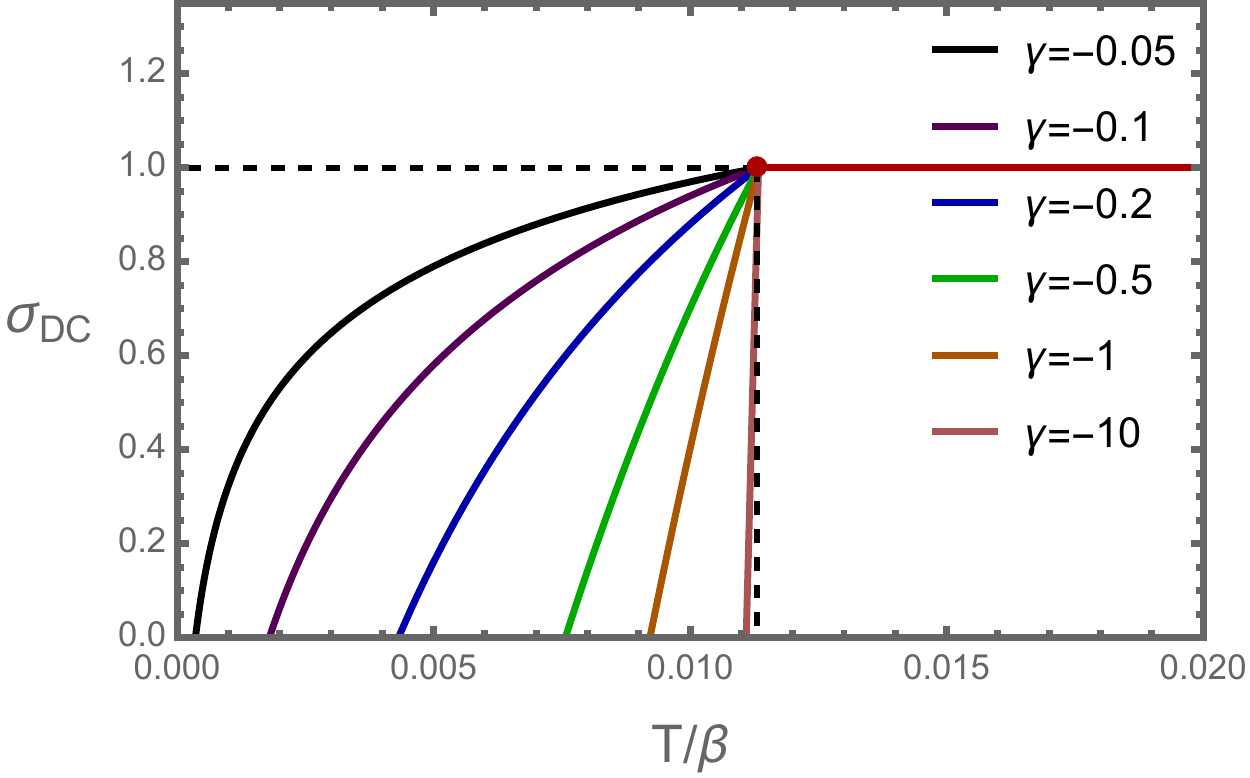}  }
   \hskip0.5cm
 \subfigure[]
   { \includegraphics[width=7cm]{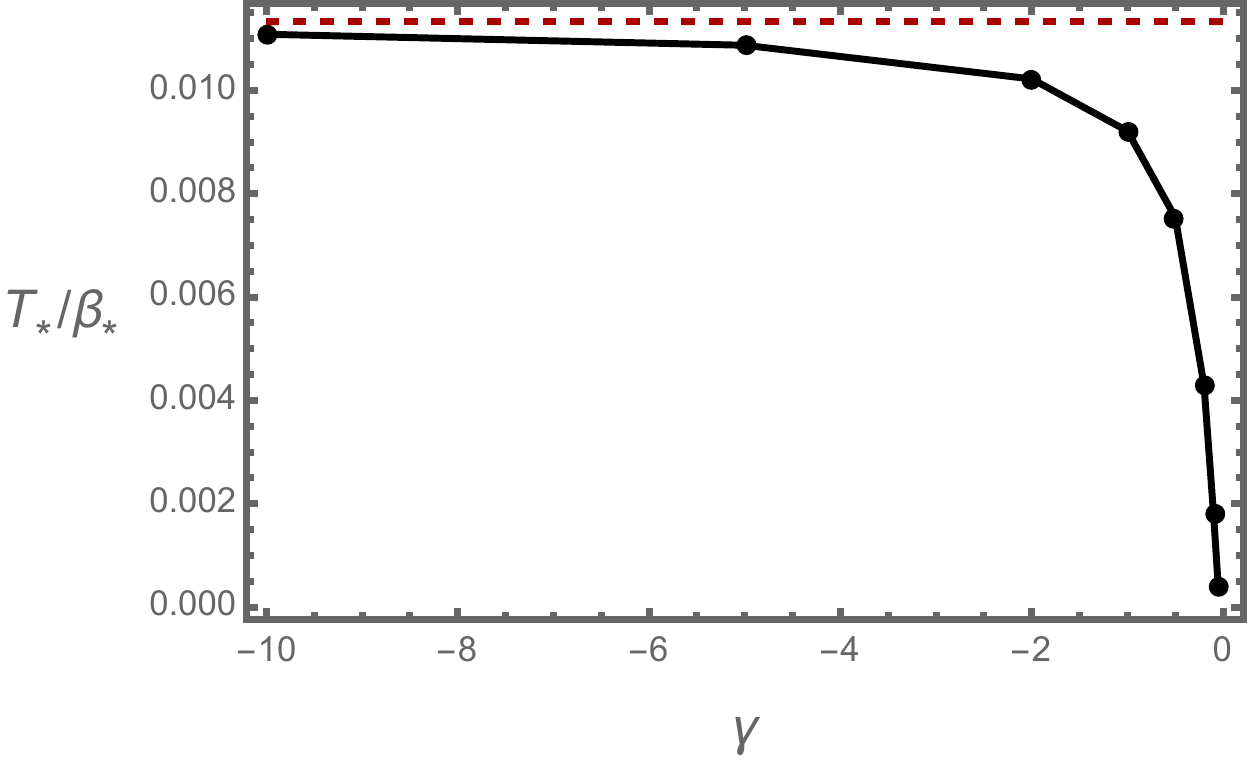}  }

\caption{(a) Temperature dependence of DC conductivity for different $\gamma $ with ${\cal Q}=0$. (b) $\gamma$ dependence of $(T/\beta)_{*}$. Red dashed line denotes to $(T/\beta)_C$.}
\label{fig:DC1}
\end{center}
\end{figure}

Figure \ref{fig:DC1} (a) shows $\gamma$ dependence of DC conductivity. Horizontal and vertical dashed lines denote DC conductivity in the limit of $|\gamma| \rightarrow 0$ and $|\gamma| \rightarrow \infty$ respectively. In the absence of $\gamma$ coupling, DC conductivity is $1$ in all temperatures, so it doesn't show any gap structure. But there still exists a phase transition from the hairy black brane to the RN-AdS black brane at $(T/\beta)_C$ that is the red dot in the figure.   The $\gamma$ dependence of $(T/\beta)_*$ is drawn in Figure \ref{fig:DC1} (b). The gap scale $(T/\beta)_*$ seems to approach to $(T/\beta)_C$ in the limit of $\gamma \rightarrow \infty$. In this limit, the gap fully occupies the insulating region completely, so only the unstable region and metallic phases exist.

\begin{figure}[ht!]
\begin{center}
   { \includegraphics[width=7cm]{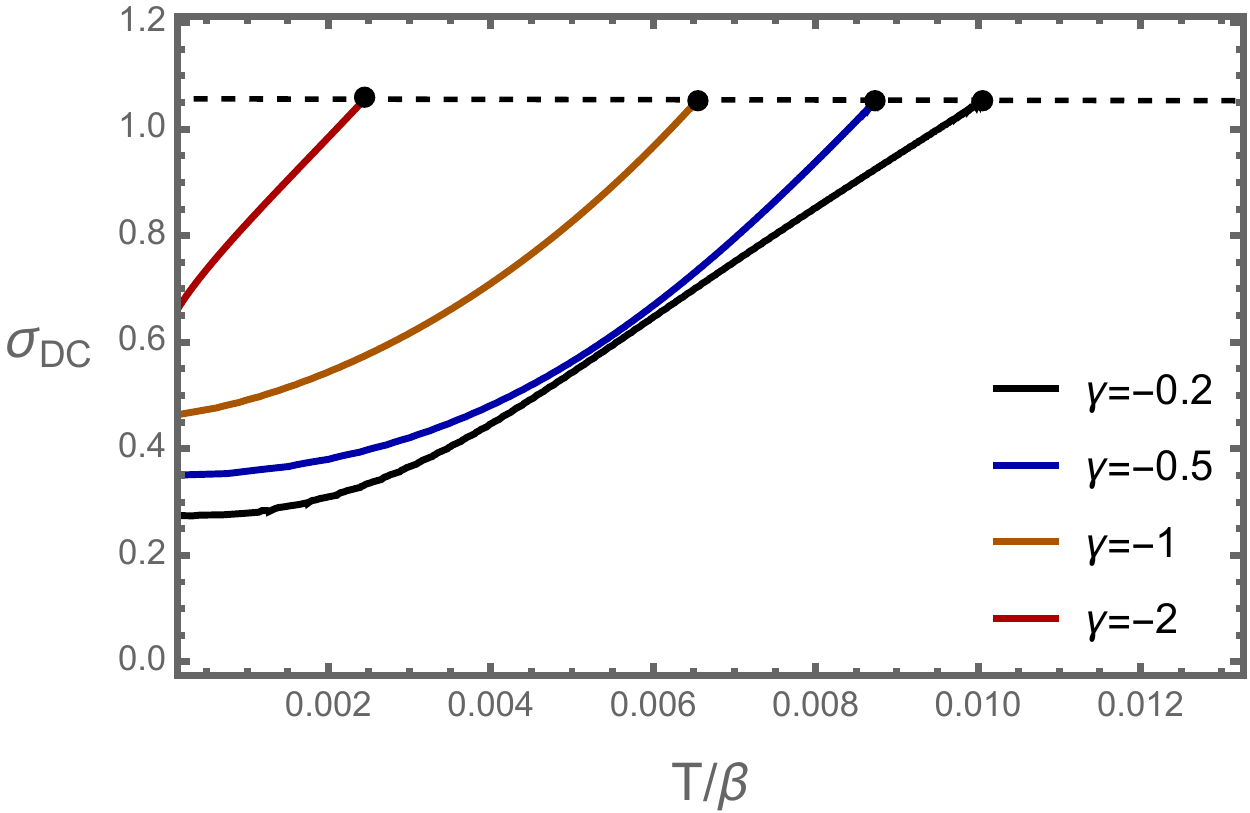}  }

\caption{Temperature dependence of DC conductivity in hairy black brane phase for different $\gamma$. Black dots denote the phase transition point to RN black brane. }
\label{fig:DC5}
\end{center}
\end{figure}

Figure \ref{fig:DC5} is a temperature dependence of DC conductivity in the insulating phase for different $\gamma$ interactions with fixed ${\cal Q}/\beta^2$. As the $\gamma$ interaction increases, the transition temperature to the metallic phase decreases. This observation is consistent with Figure \ref{fig:g2PT}.

\begin{figure}[ht!]
\begin{center}
   { \includegraphics[width=6.5cm]{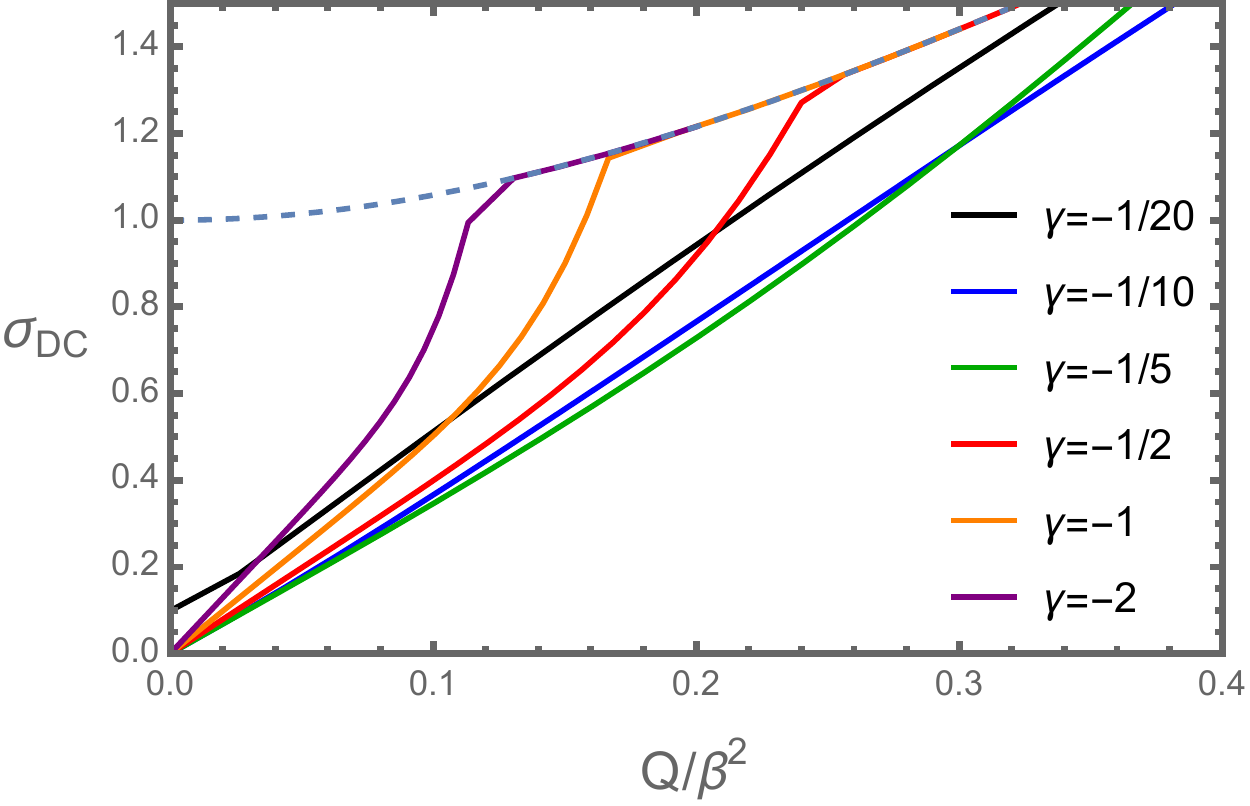}  }

\caption{The charge density dependence of DC conductivity at zero temperature for different $\gamma$ interactions.}
\label{fig:DC6}
\end{center}
\end{figure}

Figure \ref{fig:DC6} shows charge density dependence of DC conductivity at zero temperature for different $\gamma$ interactions. The dashed line denotes DC conductivity of the metallic phase which has quadratic behavior of charge carrier density.  In the figure, the decreasing ratio increases as the $\gamma$ interaction becomes strong.

\section{Quantum phase transition}\label{QCP}

This section shows that our system undergoes a phase transition at zero temperature. This phenomenon can be regarded as a `quantum phase transition. To show this, we find a parameter region where the black brane does not carry any hairy configuration at zero temperature through the BF bound argument.

In order to consider the BF bound argument, we will take the probe approximation. The equation of motion for the real scalar in the probe limit is
\begin{align}
\left(\nabla^2 -m^2 -\frac{1}{2} \gamma F^2  \right)\phi =0~.
\end{align}
The background metric is the RN-AdS with the linear axion field given by
\begin{align}
U(r) = \frac{r^2}{L^2} - \frac{\beta^2 L^2}{2} -\frac{M L^2}{r}+\frac{\mathcal{Q}^2 L^6}{4 r^2}~,~w(r) =0~,~A = \left(\mu L - \frac{\mathcal{Q} L^3}{r}\right)dt\,.
\end{align}
Since we consider a `quantum phase transition' described by a hairy black brane, the extremal black brane is suitable as the background geometry. The corresponding metric function is
\begin{align}
U(r) = U(r) = \frac{\left(r-r_h\right)^2}{L^2} \left( 1+\frac{2r_h}{r}+\frac{\mathcal{Q}L^8}{4 r^2 r_h^2} \right)~,
\end{align}
where we use a zero temperature condition $\left(\beta^2 = \frac{6 r_h^2}{L^4}-\frac{\mathcal{Q}^2 L^4}{2 r_h^2}\right)$. Thus our task is to find the scalar configuration in this background. The scalar extends from the horizon to the boundary of the extremal black brane.

The near horizon geometry of this metric is $AdS_2\times\mathbb{R}^2$. Since the BF bound argument depends on the radius of AdS space, we need to know the effective radius of $AdS_2$. The effective AdS radius is given as follows:
\begin{align}
L_{\text{eff}}^2=\frac{L^2}{3 + \frac{\mathcal{Q}^2 L^8}{4  r_h^4} }=\frac{L^2}{6}\left( 1 +\frac{\beta^2}{\sqrt{12\mathcal{Q}^2 + \beta^4}}\right)~.
\end{align}

The scalar field configuration as the lowest excitation in this probe limit is described by $\phi = \mathcal{R}(r)e^{-i\omega t}$. Then, the hairy configuration near horizon is effectively a two-dimensional probe in the $AdS_2$. In general, a scalar field has a mass bound dubbed BF bound. Only for the following case, the scalar field is stable in $AdS_{d+1}$ with the radius $L$:
\begin{align}\label{BF bound}
m^2 \geq -\frac{d^2}{4 L^2}~.
\end{align}
Near horizon, the scalar field under consideration has an effective mass given by
\begin{align}
m_{\text{eff}}^2=\left(m^2 + \frac{1}{2}\gamma F^2\right)_{r=r_h}=m^2 -\gamma\frac{\mathcal{Q}^2 L^6}{r_h^4}~.
\end{align}
Thus (\ref{BF bound}) implies that there is no hairy configuration for the following parameter region:
\begin{align}\label{instability}
m_{\text{eff}}^2\, L_{\text{eff}}^2 = \frac{1}{6}\left(1+ \frac{1}{\sqrt{1+12 {Q}^2/\beta^4}}\right)\left({m}^2L^2 - \gamma \,\frac{12^2 Q^2/\beta^4}{\left(1 +\sqrt{1 + 12 Q^2/\beta^4}\right)^2} \right)>-1/4~.
\end{align}
In particular, for the case with $m^2 = -2/L^2$ and $\gamma =-0.2$, the above condition becomes $\mathcal{Q}/\beta^2 \geq 5\sqrt{21}/38\sim0.60297$. So we may say that geometries in this parameter region at $T=0$ is a non-hairy black brane.

\begin{figure}[ht!]
\begin{center}
\subfigure[]
   { \includegraphics[width=6.5cm]{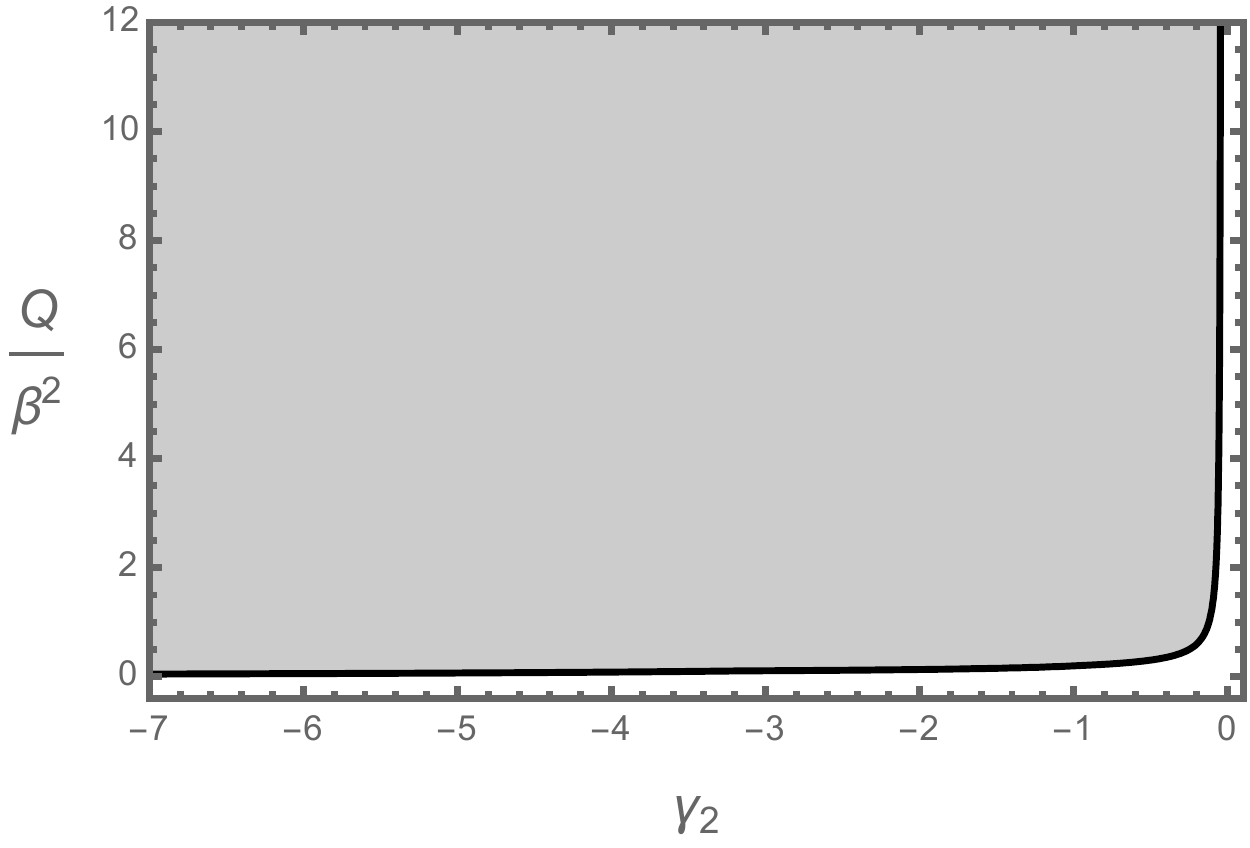}  } 
\subfigure[]
   { \includegraphics[width=6.5cm]{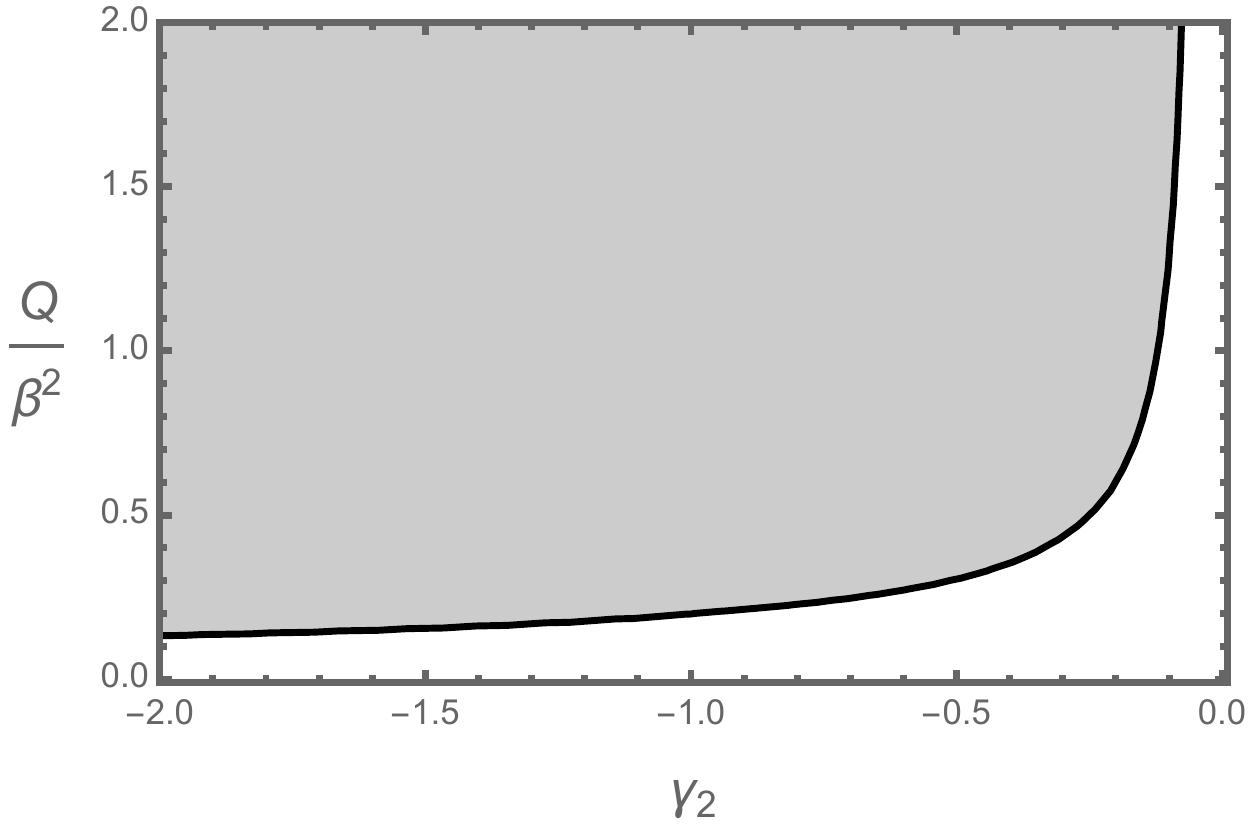}  }
    
\caption{The gray region denotes the parameter space where any hairy configuration is not allowed. See Figure \ref{fig:g2PT}. (b) lies in the white region. }
\label{fig:Extremal Region}
\end{center}
\end{figure}

By this reasoning, we claim that the `quantum phase transition' really occurs in this system. For more general cases, we plot the non-hairy black brane region where the hairy configuration is not allowed. See Figure \ref{fig:Extremal Region}. The result covers the upper region of Figure \ref{fig:g2PT}(b). Thus this BF bound argument provides the necessary condition successfully for the phase transition. Notably, this argument is independent of the size of $r_h$. In zero temperature limit, the typical behavior of the hairy black brane requires vanishing $r_h$. This limit of a different model is investigated in \cite{Horowitz:2009ij}. It would be interesting to study the zero-temperature hairy configuration in this model. We leave it as a possible future study.

\section{Scaling behavior of the resistivity}\label{scaling}
\begin{figure}[ht!]
\begin{center}
   { \includegraphics[width=8.4cm]{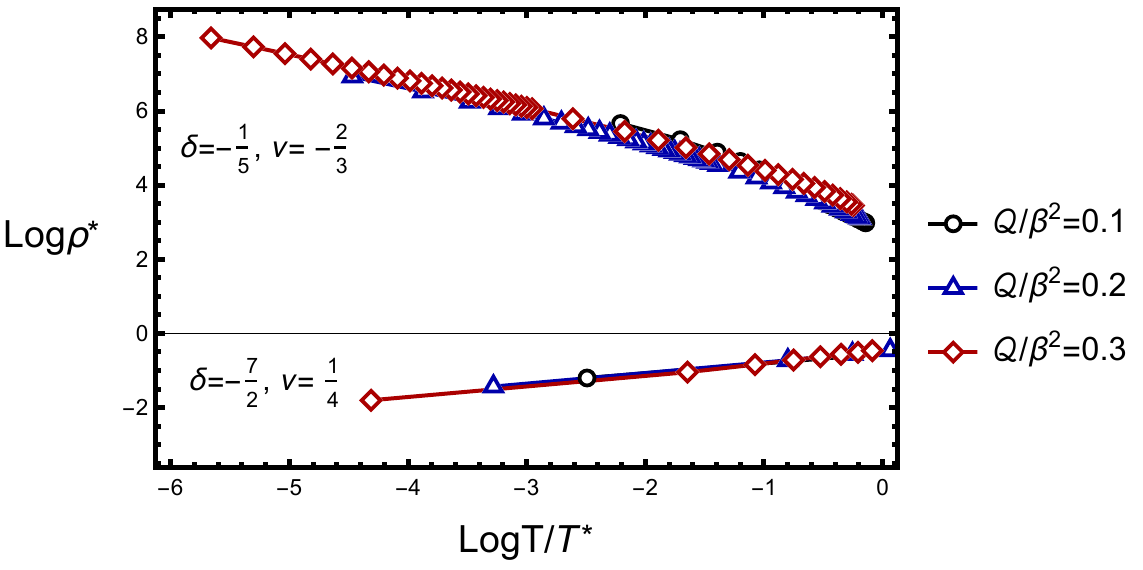}  }

\caption{Scaling behavior of DC conductivity.}
\label{fig:scaling}
\end{center}
\end{figure}

In this section, we discuss the scaling behavior of resistivity. The several cases of two-dimensional electron systems showing metal-insulator transition have interesting scaling behavior\cite{popovic1997metal,kravchenko2003metal}. The scaling property can be obtained by rescaling temperature and resistivity using critical charge density(${\cal Q}_C$) where `quantum phase transition' appears and critical temperature($T_C$) for metal-insulator transition as follows:
\begin{align}
T^* = T_C |{\cal Q} -{\cal Q}_C|^{\delta}\,,~~~~~~\rho^* = \rho T^{\nu}\,,
\end{align}
where all quantities are scaled by impurity density $\beta$ as used in the paper. Figure \ref{fig:scaling} shows the scaling behavior of resistivity for each phase. In the figure, the upper lines are resistivity in the insulating phase which decreases with temperature and the lower lines are resistivity in conducting phase which increases as the temperature increases. If we choose $(\delta,~\nu)$ to be $ (-1/5, ~-2/3)$, all resistivity lines in the insulating phase are on top of each other. With the value of $(-7/2,~1/4)$, all the resistivity lines in conducting phase are overlapped. This scaling behavior appears to wide range around the `quantum phase transition' point.

The results are different from the experimental data of metal-insulator transition materials in \cite{popovic1997metal,kravchenko2003metal}. We speculate that the reason comes from the absence of critical temperature in the experiment.

\end{appendices}

\section*{Acknowledgments}

We thank Ki-Seok Kim for helpful discussions on scaling behavior. Y. Ahn thanks to Matteo Baggioli and Hyun-Sik Jeong for helpful discussion.  Y. Seo was supported by Mid-career Researcher Program through NRF grant No. NRF-2022R1A2C1010756. K.-Y. Kim was supported by NRF funded by the Ministry of Science, ICT $\&$ Future Planning (NRF- 2021R1A2C1006791) and the GIST Research Institute(GRI) grant funded by the GIST in 2022. S. J. Sin was supported by Mid-career Researcher Program through the NRF grant No. NRF-2016R1A2B3007687. K. K. Kim was supported by Mid-career Researcher Program through NRF grant No. NRF-2019R1A2C1007396.

\end{document}